\pdfoutput=1
\documentclass[12pt]{article}

\usepackage[pdftex]{graphicx}
\usepackage{graphicx}
\usepackage{colordvi}
\usepackage{fancybox}
\usepackage{cancel}
\usepackage{amsmath}
\usepackage{amsfonts}
\usepackage{slashed}
\usepackage{amssymb}
\usepackage{caption}
\usepackage{appendix}
\usepackage{hyperref} 
\usepackage{multirow}
\hypersetup{colorlinks=true, linkcolor=black, urlcolor=blue}
\usepackage{float}
\usepackage{cite}
\usepackage{mathtools}
\usepackage{bbm}

\DeclarePairedDelimiter\floor{\lfloor}{\rfloor}

\begin{document}

\newcommand{\ba}[1]{\begin{array}{#1}} \newcommand{\ea}{\end{array}}

\numberwithin{equation}{section}


\def\Journal#1#2#3#4{{#1} {\bf #2}, #3 (#4)}

\def\NCA{\em Nuovo Cimento}
\def\NIM{\em Nucl. Instrum. Methods}
\def\NIMA{{\em Nucl. Instrum. Methods} A}
\def\NPB{{\em Nucl. Phys.} B}
\def\PLB{{\em Phys. Lett.}  B}
\def\PRL{\em Phys. Rev. Lett.}
\def\PRD{{\em Phys. Rev.} D}
\def\ZPC{{\em Z. Phys.} C}

\def\st{\scriptstyle}
\def\sst{\scriptscriptstyle}
\def\mco{\multicolumn}
\def\epp{\epsilon^{\prime}}
\def\vep{\varepsilon}
\def\ra{\rightarrow}
\def\ppg{\pi^+\pi^-\gamma}
\def\vp{{\bf p}}
\def\ko{K^0}
\def\kb{\bar{K^0}}
\def\al{\alpha}
\def\ab{\bar{\alpha}}

\def\np{Nucl. Phys. {\bf B}}\def\pl{Phys. Lett. {\bf B}}
\def\mpl{Mod. Phys. {\bf A}}\def\ijmp{Int. J. Mod. Phys. {\bf A}}
\def\cmp{Comm. Math. Phys.}\def\prd{Phys. Rev. {\bf D}}

\def\oa{\bigcirc\!\!\!\! a}
\def\ob{\bigcirc\!\!\!\! b}
\def\oc{\bigcirc\!\!\!\! c}
\def\oi{\bigcirc\!\!\!\! i}
\def\oj{\bigcirc\!\!\!\! j}
\def\ok{\bigcirc\!\!\!\! k}
\def\ve{\vec e}\def\vk{\vec k}\def\vn{\vec n}\def\vp{\vec p}
\def\vr{\vec r}\def\vs{\vec s}\def\vt{\vec t}\def\vu{\vec u}
\def\vv{\vec v}\def\vx{\vec x}\def\vy{\vec y}\def\vz{\vec z}

\def\ve{\vec e}\def\vk{\vec k}\def\vn{\vec n}\def\vp{\vec p}
\def\vr{\vec r}\def\vs{\vec s}\def\vt{\vec t}\def\vu{\vec u}
\def\vv{\vec v}\def\vx{\vec x}\def\vy{\vec y}\def\vz{\vec z}

\newcommand{\AdS}{\mathrm{AdS}}
\newcommand{\dd}{\mathrm{d}}
\newcommand{\eee}{\mathrm{e}}
\newcommand{\sgn}{\mathop{\mathrm{sgn}}}

\def\a{\alpha}
\def\b{\beta}
\def\g{\gamma}

\newcommand\lsim{\mathrel{\rlap{\lower4pt\hbox{\hskip1pt$\sim$}}
    \raise1pt\hbox{$<$}}}
\newcommand\gsim{\mathrel{\rlap{\lower4pt\hbox{\hskip1pt$\sim$}}
    \raise1pt\hbox{$>$}}}

\newcommand{\beq}{\begin{equation}}
\newcommand{\eeq}{\end{equation}}
\newcommand{\bea}{\begin{eqnarray}}
\newcommand{\eea}{\end{eqnarray}}
\newcommand{\bem}{\begin{pmatrix}}
\newcommand{\eem}{\end{pmatrix}}
\newcommand{\noi}{\noindent}
\newcommand{\non}{\nonumber}
\newcommand{\rdec}{\color{red}}
\newcommand{\plav}{\color{blue}}
\newcommand{\bet}{\begin{itemize}}
\newcommand{\eet}{\end{itemize}}
\newcommand{\ben}{\begin{enumerate}}
\newcommand{\een}{\end{enumerate}}


\begin{flushright}
December, 2022
\end{flushright}

\bigskip

\begin{center}

{\Large\bf  Holographic thermal propagator for arbitrary scale dimensions
 }
\vspace{1cm}

\centerline{
Borut Bajc$^{a}$\footnote{borut.bajc@ijs.si} 
and 
Adri\'{a}n R. Lugo$^{b,c}$\footnote{lugo@fisica.unlp.edu.ar}
}
\vspace{0.5cm}
\centerline{
$^{a}$ {\it\small J.\ Stefan Institute, 1000 Ljubljana, Slovenia}
}

\centerline{
$^{b}$ {\it\small Instituto de F\'isica La Plata - CONICET,  Diagonal 113 y 63, 1900 La Plata, Argentina}
}
\centerline{
$^{c}$ {\it\small Departamento de F\'\i sica, Facultad de Ciencias Exactas, Universidad Nacional de La Plata}
}
\centerline{
{\it\small  C.C. 67, 1900 La Plata, Argentina}
}

\end{center}

\bigskip

\begin{abstract}
Using the AdS/CFT correspondence we model the behaviour of the two-point correlator 
of an operator with arbitrary scale dimension $\Delta$ in arbitrary spacetime dimension $d$ for 
small but non-zero temperature. The obtained propagator coincides in the low temperature regime 
with the known result for $d=4$ for large $\Delta$ at the order $T^d$ as well as with the $T^d$ 
and $T^{2d}$ terms of the exact all order result for $d=2$. Furthermore, for arbitrary $d$ we 
explicitly write down the expression for the order $T^{d}$ of the propagator for arbitrary $\Delta$, 
and present a conjecture for the order $T^{2d}$ in the large $\Delta$ limit.

\end{abstract}


\tableofcontents

\section{Introduction}

Correlation functions of  operators in conformal field theories are 
constrained by symmetry. For example, the two-point correlation function in 
$d$-dimensional Euclidean momentum space of a spin zero primary operator 
with scale dimension $\Delta=d/2+\nu$ is given by

\beq
G_2^{(0)}(k)\propto k^{2\nu}\quad,\quad k^2=\omega^2+\vec{q}^{\,2}\quad,
\eeq

\noi
where we  split the time and space part of  the $d$-dimensional vector $k^\mu=(\omega,\vec{q})$ 
for later use.

A bit less obvious is what happens when in such a theory we introduce a non-zero temperature, i.e. 
in the spacetime $S_{1/T}^1\times \mathbb{R}^{d-1}$. Conformal symmetry is now broken by $T\ne0$ 
and correlators get, as we will see, corrections of order ${\cal O}((T/k)^d)$. There have been recently 
two important developments in this direction. First, the problem has been attacked 
in the limit of $\vec{q}=0$ for the absorption cross section by black branes \cite{Policastro:2001yb} or 
using the AdS/CFT correspondence \cite{Maldacena:1997re,Gubser:1998bc,Witten:1998qj} 
for large scale dimensions $\Delta\gg1$ 
\cite{Fitzpatrick:2019zqz,Rodriguez-Gomez:2021pfh,Rodriguez-Gomez:2021mkk}. This last limit was needed to be able to apply the geodesic 
approximation in the background geometry of a black brane in AdS, while the leading order has been given 
for general $\Delta$ recently  in terms of the ambient space formalism \cite{Parisini:2022wkb}. Second, by using the solution 
to the connection problem of the Heun equation \cite{CarneirodaCunha:2015qln, Bonelli:2022ten} 
a complete analytic result for the thermal two-point correlation function has been found recently in 
\cite{Dodelson:2022yvn} (with the generalisation to non-zero chemical potential in \cite{Bhatta:2022wga}).

These results are however not completely satisfactory. First, large scale dimensions are not necessarily 
realistic, so results for smaller values are welcome. Second, although the solution found in \cite{Dodelson:2022yvn} 
is analytic, it is very hard to make it work properly in the small temperature 
expansion of the black brane case, mainly because an infinite number of terms  contribute and the 
resummation is far from obvious. 

In this work we want to improve the state of the art calculating 
the first two corrections (the $(T/k)^d$ and the $(T/k)^{2d}$ ones) to the zero temperature propagator 
for a completely generic scale dimension $\Delta$ and arbitrary spacetime dimension $d$. 
In some sense this paper is a continuation of \cite{Bajc:2012vk}, although it uses a different approach. 
The method for calculating the propagator we use in this paper is straightforward: 
we rewrite the usual second order ordinary differential equation (ODE) for a perturbation in 
the black brane AdS background as a matrix linear first order ODE, expand it in powers 
of the temperature and 
solve it by explicit integration. When we will not be able to explicitly and analytically compute the integrals 
involved, we will evaluate them numerically. As we will show, our method turns out to be 
tougher and tougher as the scale dimension increases, so it is particularly useful especially for 
low $\Delta$ and so it is complementary to the large $\Delta$ calculation of 
\cite{Fitzpatrick:2019zqz,Rodriguez-Gomez:2021pfh,Rodriguez-Gomez:2021mkk}.

\section{The propagator from the AdS/CFT approach}

As we mentioned in the introduction we will model our CFT primary operator with scale dimension 
$\Delta$ by a scalar field in the AdS black brane background. The equation to solve in momentum 
space for the scalar field perturbation 

\beq
\eta(z)=\int d^{d}x\, e^{-ix_\mu k^\mu}\phi(x,z)
\eeq

\noi
is thus (see for example \cite{Aharony:1999ti} and references therein)

\beq\label{eq:etaeq}
\left(f(z/z_h)\;z\;\frac{d}{dz}\;z\frac{d}{dz} -d\;z\; \frac{d}{dz} - z^2\;\left( k^2 +\omega^2\;\frac{1-f(z/z_h)}{f(z/z_h)}\right) - \Delta(\Delta-d)\right)\,\eta(z) = 0\quad,
\eeq
where 

\beq
f(x)\equiv 1-x^d
\eeq

\noi
with the AdS radius $L=1$ and the horizon radius (black brane position)  
expressed in terms of the spacetime dimension $d$ and temperature $T$ through

\beq
z_h=\frac{d}{4\pi T}\quad.
\eeq

Our choice of (Euclidean) boundary conditions for the two linearly independent solutions $\eta_\pm(z)$ are

\bea
\label{boundaryeta}
z\rightarrow 0^+&:&\eta_\pm(z)\quad\longrightarrow z^{\frac{d}{2}\pm\nu}\quad,\non\\
z\rightarrow z_h&:&|\eta_-(z)|<\infty\quad,
\eea

\noi
where

\beq
\label{nu}
\nu=\Delta-d/2
\eeq

\noi
is taken to be positive with non integer $2\nu$ all along the paper.

If we write

\bea\label{eq:etah}
\eta(z) = \frac{z^\frac{d-1}{2}}{f(z/z_h)^\frac{1}{2}}\;h(z)\quad,
\eea

\noi
then \eqref{eq:etaeq} implies that $h(z)$ obeys

\beq\label{eq:h}
\left(\frac{d^2}{dz^2} - U(z)\right)\, h(z) = 0\quad,
\eeq

\noi
where

\beq
\label{eq:U}
U(z) = -\frac{1}{4\,z^2} + \frac{d^2}{4}\,\frac{2\,f(z/z_h)-1}{z^2\,f(z/z_h)^2} + \frac{\Delta(\Delta-d)}{z^2\,f(z/z_h)} + 
\frac{ k^2}{f(z/z_h)} 
+ \omega^2\,\frac{1-f(z/z_h)}{f(z/z_h)^2}\quad.
\eeq

Let  $h_\pm(z)$ be a basis of solutions to \eqref{eq:h}. 
The general solution can be written as

\beq\label{eq:hxact}
h(z) = C_+\,h_+(z) + C_-\,h_-(z)\quad,
\eeq

\noi
where $C_\pm$ are arbitrary constants. 
Taking into account the derivative of \eqref{eq:hxact}, 
we can write,

\beq\label{eq:hh'exact}
\left(\begin{array}{c}h(z)\\ h'(z)\end{array}\right)
= {\bf w}\left(h_+,h_-;z\right)\;
\left(\begin{array}{c}C_+\\ C_-\end{array}\right)\quad,
\eeq

\noi
where 

\beq\label{eq:hh'wC}                   
{\bf w}\left(f,g;z\right)\equiv 
\left(\begin{array}{cc}
f(z)&g(z)\\ f'(z)&g'(z)\end{array}\right)
\eeq

\noi
is the wronskian matrix of $f$ and $g$ at the point $z$. 
For any basis of solutions of \eqref{eq:h} its determinant, i.e. the wronskian, is a non-zero constant. 
From \eqref{eq:hh'exact}, our problem reduces to the computation of the wronskian matrix of a given basis. 
According to (\ref{boundaryeta}) the (Euclidean) boundary conditions are

\bea
z\rightarrow 0^+&:&h_\pm(z)\quad\longrightarrow z^{\frac{1}{2}\pm\nu}\quad,\non\\
\label{boundaryh}
z\rightarrow z_h&:&|h_-(z)|=0\quad.
\eea

\noi
In this way, the wronskian results: 
$\det({\bf w}\left(h_+,h_-;z\right))= -2\,\nu$.

We can recast the second order equation \eqref{eq:h} as the following  first order system,

\beq
\label{eq:firstorder}
\frac{d}{dz}\left(
\begin{array}{c}h(z)\\ h'(z)\end{array}\right)
= {\bf A}(z)\;
\left(\begin{array}{c}h(z)\\ h'(z)\end{array}\right)\quad,\quad 
{\bf A}(z)\equiv
\left(\begin{array}{cc}	0&1\\ U(z)&0\end{array}\right)\quad,
\eeq

\noi
or, for the wronskian matrix 

\beq\label{eq:Weq}
\frac{d}{dz} {\bf w}\left(h_+,h_-;z\right) = 
{\bf A}(z)\;{\bf w}\left(h_+,h_-;z\right)\quad.
\eeq

The potential ${\bf A}(z)$ can be split into a zero-temperature and a temperature-dependent part as

\beq\label{eq:A=A0+AT}
{\bf A}(z)= {\bf A}^{(0)}(z) + {\bf A}_T(z)\quad,
\eeq

\noi
where the $T=0$ matrix is

\beq
\label{eq:A0}
{\bf A}^{(0)}(z)= k^2\;
\left(\begin{array}{cc}	0 & 1\\  1+ \frac{\nu^2 -\frac{1}{4}}{q^2}&0\end{array}\right)\quad,
\eeq

\noi
with

\beq
q\equiv kz\quad,
\eeq

\noi
while that the temperature-dependent part can be expanded in  powers of  $\left(T/k\right)^d$ as

\beq
\label{eq:AT}
{\bf A_T(z)} =  k^2\; \sigma_-u_T(q)\quad,\quad
u_T(q)=\sum_{m\in\mathbb N^+} \frac{ u^{(m)}(kz)}{ (kz_h)^{d\,m}}\quad,
\eeq

\noi
where

\bea
\label{eq:um}
\sigma_-&=&
\bem
0 & 0 \\
1 & 0
\eem\quad,\non\\
u^{(m)}(q)&=&A_m\, q^{d\,m-2} + B_m\, q^{d\,m}\quad,\non\\
A_m\equiv \nu^2-m\,\frac{d^2}{4}&,&
B_m\equiv 1 + m\,(\omega/k)^2\quad.
\eea

\subsection{The $T=0$ case}

At zero temperature \eqref{eq:Weq} reads,
\beq\label{eq:W0eq}
\frac{d}{dz} {\bf w}\left(h^{(0)}
_+,h^{(0)}_-;z\right)= 
{\bf A}^{(0)}(z)\;{\bf w}\left(h^{(0)}_+,h^{(0)}_-;z\right)\quad.
\eeq

\noi
The solutions are $z^\frac{1}{2}$ times Bessel functions of order $\nu$. 
A summary of their properties and useful expansions is given in appendix \eqref{app:someuseful}. 
The basis \eqref{boundaryh} is given by
\bea
h^{(0)}_+(z) &=&\frac{2^\nu\,\Gamma(1+\nu)}{ k^\nu}\;z^\frac{1}{2}\;I_\nu(q) \quad,\\
h^{(0)}_-(z)&=&
\frac{2^{1-\nu}\, k^\nu}{\Gamma(\nu)}\;z^\frac{1}{2}\;K_\nu(q)\quad.
\eea

\noi
According to the AdS/CFT 
recipe the propagator is defined through the limit
\beq\label{eq:propdefT=0}
h_-(z)\quad\stackrel{z\rightarrow 0^+}{\longrightarrow}\quad z^{\frac{1}{2}-\nu}\;(1+\dots) +
 G_2(k)\;z^{\frac{1}{2}+\nu}\;(1+\dots)
\eeq
to be in the leading order
\beq\label{eq:propT=0}
 G_2^{(0)}(k) = \frac{\Gamma(-\nu)}{\Gamma(\nu)}\;
\left(\frac{k}{2}\right)^{2\,\nu} \quad.
\eeq

\subsection{The $T\neq0$ case}

Going to the general case, if we make the ansatz for any $T$,
\beq\label{eq:WW0WT}
{\bf w}\left(h_+,h_-;z\right) =
{\bf w}\left(h^{(0)}_+,h^{(0)}_-;z\right)\;
\left( {\bf 1} + {\bf w}^{(T)}(q)\right)\quad,
\eeq

\noi
then from \eqref{eq:Weq} and \eqref{eq:W0eq} we get for 
${\bf w}^{(T)}(q)$

\beq\label{eq:WTeq}
\frac{d}{dq} {\bf w}^{(T)}(q) = 
q\;u_T(q)\;{\bf b}(q)\;\left( {\bf 1} + {\bf w}^{(T)}(q)\right)\quad,
\eeq

\noi
where $u_T(q)$ is given in \eqref{eq:AT} 
and the matrix ${\bf b}(q)$ is
\bea
\label{eq:b}
{\bf b}(q) &\equiv& z^{-1}\; {\bf w} \left(h^{(0)}_+,h^{(0)}_-;z\right)^{-1}\;\sigma_-\;
{\bf w}\left(h^{(0)}_+,h^{(0)}_-;z\right)\non\\
&=& \left(\begin{array}{cc}
I_\nu(q)\,K_\nu(q)
&\alpha\,K_\nu(q){}^2\\-\frac{1}{\alpha}\,I_\nu(q){}^2& -I_\nu(q)\,K_\nu(q)
\end{array}\right)\quad,
\eea

\noi
where

\beq
\label{alpha}
\alpha\equiv \frac{2^{1-2\nu}\, k^{2\nu}}{\Gamma(\nu)\,\Gamma(1+\nu)}
= -2\frac{\sin(\pi\nu)}{\pi}\, G_2^{(0)}(k)\quad.
\eeq

In order to solve \eqref{eq:WTeq} we expand

\beq
\label{expwT}
{\bf w}^{(T)}(q) = \sum_{m\in\mathbb N^+} 
\frac{{\bf w}^{(m)}(q)}{(kz_h)^{m\,d}}
\eeq

\noi
and get the following system of differential equations for the ${\bf w}^{(m)}(q)$'s, $m>0$ 
(we define ${\bf w}^{(0)}=\mathbb{I}$)

\beq
\label{eq:wmeq}
\frac{d}{dq} {\bf w}^{(m)}(q) =
 \sum_{k=0}^{m-1}\;
q\;u^{(m-k)}(q)\;{\bf b}(q)\;{\bf w}^{(k)}(q)\quad,
\eeq

\noi
to be solved iteratively, with the $u^{(m)}(q)$'s given in \eqref{eq:um}.
The solution is

\beq
\label{eq:wm}
{\bf w}^{(m)}(q) = {\bf w}_0^{(m)} +
\sum_{k=0}^{m-1}\;
\int^q dz\,z\;u^{(m-k)}(z)\;{\bf b}(z)\;{\bf w}^{(k)}(z)\quad.
\eeq

\noi
From \eqref{eq:WW0WT} and (\ref{eq:hh'wC}) we have

\bea
\label{eq:hpmTgral}
h_+(z) &=& \left(1+ w_{11}^{(T)}(q)\right)\;h^{(0)}_+(z) +  w_{21}^{(T)}(q)\;h^{(0)}_-(z) \quad, \cr
&&\cr
h_-(z) &=& w_{12}^{(T)}(q)\;h^{(0)}_+(z) + 
\left(1+ w_{22}^{(T)}(q)\right)\;h^{(0)}_-(z)\quad.
\eea

\noi
The integration constants present in the computation of the matrix elements 
$w_{ij}^{(T)}(q)$ are fixed by (\ref{boundaryh}), i.e. for $z\to0$

\beq
\label{h0}
z\rightarrow 0^+\quad:\quad h_\pm(z)\quad\longrightarrow z^{\frac{1}{2}\pm\nu}\quad.
\eeq

\noi
Since expanding everything around $T=0$ is equivalent to 
expanding everything around $z_h=\infty$, we replace the last constraint of  
(\ref{boundaryh}) with

\beq
\label{hinf}
z\rightarrow \infty\quad:\quad h_-(z)\longrightarrow 0\quad.
\eeq

Near $z=0$, 

\bea
h_+(z) &=& \left(1+ w_{11}^{(T)}(q)\right)\;
z^{\frac{1}{2}+\nu}\,(1+{\cal O}(q))\cr
&+& w_{21}^{(T)}(q)\;z^{\frac{1}{2}-\nu}\,
\left(1+{\cal O}(q) +  G_2^{(0)}(k)\,z^{2\,\nu}\,(1+{\cal O}(q))\right)\quad,\cr 
h_-(z) &=& w_{12}^{(T)}(q)\;z^{\frac{1}{2}+\nu}\,(1+{\cal O}(q))\cr
&+& \left(1+ w_{22}^{(T)}(q)\right)\,z^{\frac{1}{2}-\nu}\,
\left(1+{\cal O}(q) +  G_2^{(0)}(k)\,z^{2\,\nu}\,(1+{\cal O}(q))\right)\quad.
\eea

\noi
Conditions \eqref{h0} are fulfilled if

\beq
\label{eq:wijcond}
0 = w_{11}^{(T)}(0)  = w_{22}^{(T)}(0) = \left.q^{-2\,\nu}\,w_{21}^{(T)}(q)\right|_{q\rightarrow 0} 
=\left.q^{2\,\nu}\,w_{12}^{(T)}(q)\right|_{q\rightarrow 0}\quad.
\eeq

The first three conditions imply that $w_{11,21,22}^{(T)}(0)=0$, 
so they determine the respective integration constants $w_{0\,ij}^{(T)}$ 
in \eqref{eq:wm}. Since in general up to a constant

\bea
\label{wij0}
z\to0&:& w_{ij}^{(m)}(z)\sim z^{md+(i-j)2\nu}\quad,\\
z\to\infty&:& w_{ij}^{(m)}(z)\sim z^{md+1-|i-j|}e^{(i-j)2z}\quad,
\eea

\noi
all $(ij)$ except eventually $(12)$ are well behaved close to $z=0$. On the other side, 
close to $z=\infty$ it is exactly the $(12)$ component which is well behaved. We 
can thus rewrite \eqref{eq:wm} incorporating \eqref{h0} as 
(let define $w_{ij}^{(0)}=\delta_{ij}$) 

\beq
\label{wij}
w_{ij}^{(m)}(q) = 
\sum_{k=0}^{m-1}\;
\int_0^q dz\,z\;u^{(m-k)}(z)\;\left({\bf b}(z)\;{\bf w}^{(k)}(z)\right)_{ij}
\eeq

\noi
for $(ij)=(11), (21), (22)$, while \eqref{hinf} implies

\beq
\label{w120}
w_{12}^{(m)}(q) = 
-\sum_{k=0}^{m-1}\;
\int_q^{\infty} dz\,z\;u^{(m-k)}(z)\;\left({\bf b}(z)\;{\bf w}^{(k)}(z)\right)_{12}\quad.
\eeq

Taking into account that the behavior of a solution is always of the form: 
$a_- z^{\frac{1}{2}-\nu}\, (1+{\cal O}(z)) + a_+ z^{\frac{1}{2}+\nu}\, (1+{\cal O}(z))$, 
we can write the following behaviour for $w_{12}^{(T)}(q)$ near $q=0$,

\beq
\label{eq:wT12q=0}
w_{12}^{(T)}(q)\quad \stackrel{q\rightarrow 0}{\longrightarrow}\quad 
 G_2^{(0)}(k)\, g^{(T)}(k) + a\,q^{-2\,\nu + d}\,\left(1+{\cal O}(q)\right)\quad,
\eeq 

\noi
where $ g^{(T)}(k)$ is some free constant. Then the AdS/CFT recipe (\ref{eq:propdefT=0}) 
gives the propagator in the form

\beq
\label{eq:G2gral}
 G_2(k) =  G_2^{(0)}(k)\,\left(1+ g^{(T)}(k)\right)\quad.
\eeq

\noi
Similarly as ${\bf w}^{(T)}(k)$ in \eqref{expwT} also $g^{(T)}(k)$ gets expanded:
 
\beq
g^{(T)}(k)=\sum_{m=1}^\infty \frac{g_m(\omega/k)}{(kz_h)^{md}}\quad.
\eeq

Thus, to get the finite temperature correction to the propagator one needs to 
extract the constant part of the expansion of $w_{12}^{(T)}(q)$ for $q\to0$. 
If the integral \eqref{w120} converges, then the constant part of the expansion 
(\ref{eq:wT12q=0}) is simply

\beq
\label{gTkoncen}
G_2^{(0)}(k)g_m(\omega/k)=-\sum_{k=0}^{m-1}\;
\int_0^{\infty} dz\,z\;u^{(m-k)}(z)\;\left({\bf b}(z)\;{\bf w}^{(k)}(z)\right)_{12}\quad.
\eeq

\noi
However, for $2\nu-md>0$ the integral \eqref{w120} diverges. In this case, to extract the 
constant part, one needs first to split the integrand

\beq
F(z)\equiv-\sum_{k=0}^{m-1}\,z\;u^{(m-k)}(z)\;\left({\bf b}(z)\;{\bf w}^{(k)}(z)\right)_{12}\to
\sum_{\alpha>0}\frac{F_{-\alpha}}{z^{\alpha+1}}+\sum_{\beta>0}F_{\beta}z^{\beta-1}
\eeq

\noi
as $z\to0$, where $\alpha,\beta$ are in general non integer real numbers. Notice that 
for finite $\nu$ there is always only a finite number of divergent terms ($\alpha$). 
In this case \eqref{gTkoncen} becomes

\beq
\label{gT}
G_2^{(0)}(k)g_m(\omega/k)=
\int_0^{\infty} dz\,\left(F(z)-\sum_{\alpha>0}\frac{F_{-\alpha}}{z^{\alpha+1}}\right)\quad,
\eeq

\noi
which renders the integration at $z=0$ finite without changing the convergence at $z=\infty$.

Of course, if one is able to compute analytically \eqref{w120}, then this is not needed, and one 
just takes the constant term in the expansion around $q\to0$ of the result as indicated in eq. 
(\ref{eq:wT12q=0}). But apart from the 
$m=1$ case we were unable to perform such an analytic integration, in which case the formula 
\eqref{gT} (which reduces to \eqref{gTkoncen} for $2\nu-md<0$) is the one we will use in the 
numerical evaluation. 

\section{First correction: $(T/k)^d$}

At first order we obtain

\beq
\label{eq:w1}
{\bf w}^{(1)}(q)= A_1
\left(\begin{array}{cc}
I_d[I_\nu\,K_\nu;q,0]
&\alpha\,I_d[K_\nu{}^2;q,\infty]
\\\\-\frac{1}{\alpha}\,I_d[I_\nu{}^2;q,0]
&-I_d[I_\nu\,K_\nu;q,0]
\end{array}\right)
+\left((A_1,d)\to(B_1,d+2)\right)\quad,
\eeq

\noi
where we  defined the integral

\beq\label{eq:Is}
I_s[f;q,q_0] \equiv \int_{q_0}^q\,dz\;z^{s-1}\;f(z)\quad.
\eeq

\noi
Although we will not use them, these integrals are explicitly computed for completeness in 
Appendix \ref{app:someuseful}.

From \eqref{eq:hpmTgral} we have in our case

\begin{align}
\label{eq:hpmTfirstorder}
h_+(z) &= \left(1+\frac{1}{( kz_h)^d}\; w_{11}^{(1)}(q)\right)\;h^{(0)}_+(z) + \frac{1}{( kz_h)^d}\; w_{21}^{(1)}(q)\;h^{(0)}_-(z) 
+ {\cal O}\left(\frac{1}{( kz_h)^{2\,d}}\right)\quad,\non\\
h_-(z) &=  \frac{1}{( kz_h)^d}\;w_{12}^{(1)}(q)\;h^{(0)}_+(z) + 
\left(1+\frac{1}{( kz_h)^d}\; w_{22}^{(1)}(q)\right)\;h^{(0)}_-(z)
+ {\cal O}\left(\frac{1}{( kz_h)^{2\,d}}\right)\quad.\cr
&&
\end{align}

Using the definitions (\ref{eq:w1}) we find the leading low $q$ expansion as

\beq
\label{eq:w1ijlowq}
w_{11}^{(1)}(q)\sim q^d\quad,\quad 
w_{22}^{(1)}(q)\sim q^d\quad,\quad 
w_{21}^{(1)}(q)\sim q^{2\,\nu+d}\quad,
\eeq

\noi 
while the only constant piece is in

\beq
\label{w12}
w_{12}^{(1)}(q)\sim G_2^{(0)}(k)\left(A_1\alpha_d+B_1\alpha_{d+2}\right)
+{\cal O}\left(q^{-2\,\nu+d}\right)\quad,
\eeq

\noi
\noi
where we have defined 

\beq
\label{alphasintegral}
\alpha_s=\frac{2\sin{(\pi\nu)}}{\pi}\int_0^\infty dzz^{s-1}K_\nu^2(z)\quad,
\eeq

\noi
which is taken at face value for $s-2\nu>0$, while for $s-2\nu<0$ it must be understood 
in the sense of (\ref{gT}), i.e. with proper subtractions. However, in both cases 
(we are interested in positive integer $s$ and $\nu>d/2$) the final result is 

 \beq
\label{alphas}
\alpha_s=\frac{\sqrt{\pi}}{2}\,
\frac{\Gamma\left(\frac{s}{2}\right)}{\Gamma\left(\frac{s+1}{2}\right)}\,
\nu\frac{\Gamma(\frac{s}{2}+\nu)\Gamma(\frac{s}{2}-\nu)}{\Gamma(1+\nu)\Gamma(1-\nu)}\quad.
\eeq

The propagator (\ref{eq:propdefT=0}) at this order is

\beq
\label{eq:G2}
 G_2(k) =  G_2^{(0)}(k)\left(1+ \frac{g_1(\omega/k)}{( kz_h)^d}\right)\quad,
\eeq

\noi
with

\bea
\label{eq:w1012}
g_1(\omega/k)&=& A_1\alpha_d+B_1\alpha_{d+2}\non\\
&=&\frac{\sqrt{\pi}}{4}\frac{\Gamma\left(\frac{d}{2}\right)}{\Gamma\left(\frac{d+3}{2}\right)}
\frac{\Gamma\left(\frac{d}{2}+1+\nu\right)\Gamma\left(\frac{d}{2}+1-\nu\right)}
{\Gamma\left(1+\nu\right)\Gamma\left(1-\nu\right)}\nu
\left(d\,\left(\frac{\omega}{k}\right)^2-1\right)\quad,
\eea

\noi
which agrees with \cite{Parisini:2022wkb}. For $d=4$ we get 

\beq\label{eq:sigmad=4}
G_2(k)/G_2^{(0)}(k)-1 = \left(\frac{\pi T}{k}\right)^4\;\frac{2}{15}\; \nu\,(\nu^2-1)\,(\nu^2-4)\,\left(4\,\left(\frac{\omega}{k}\right)^2-1\right)\quad,
\eeq

\noi
which is the Fourier transform ({\ref{eq:G1}) of the result in \cite{Rodriguez-Gomez:2021pfh} 
using $\Delta=2+\nu$ \eqref{nu}. Notice that the result here is exactly the same as for large $\Delta$. 
This peculiarity is not preserved at next order in the temperature expansion. 

Since the $d=2$ solution is known \cite{Grozdanov:2019uhi}, for even $d$ there is an even easier 
way to compute this leading order correction to the propagator, 
without explicitly calculating the integral (\ref{alphasintegral}), which we now explicitly write as a 
function of $\nu$:

\beq
\label{alpha0}
\alpha_r(\nu)=\frac{2}{\pi}\sin{(\pi\nu)}\int_{0}^\infty dzz^{r-1}K_\nu^2(z)\quad.
\eeq

\noi
Using

\beq
K_{\nu}=-K_{\nu+1}'-\frac{\nu+1}{z}K_{\nu+1}
\eeq

\noi
one can derive, applying repeatedly differentiation by parts 

\beq
\label{rekurzija}
\alpha_{r+2}(\nu)=\alpha_{r+2}(\nu+1)-\frac{r(r-2\nu-2)}{2}\alpha_{r}(\nu+1)\quad.
\eeq

Eq. (\ref{rekurzija}) means that there is a relation for the $\alpha_{d+2}$ once 
$\alpha_{d}$ is known. Since $\alpha_2$ is known exactly, $\alpha_4$ can be 
calculated. 

Take the ansatz\footnote{The constant term is missing because $g_1$ is odd under 
$\nu\to-\nu$, see section \ref{nuvminusnu}, while higher orders $c_n$ for $n\geq d$ 
turn out to be zero.}

\beq
\alpha_4(\nu)=\sum_{n=1}^{d-1}c_n\nu^n
\eeq

\noi
and plugging it into (\ref{rekurzija}) we obtain ($\alpha_2(\nu)=\nu$)

\beq
c_3\left((\nu+1)^3-\nu^3\right)+c_2\left((\nu+1)^2-\nu^2\right)+c_1\left((\nu+1)-\nu\right)=
-2\nu(\nu+1)\quad.
\eeq

Equating the same powers of $\nu$ we get

\beq
\alpha_4(\nu)=\frac{2}{3}\nu\left(1-\nu^2\right)\quad,
\eeq

\noi
which is the correct result got by direct integration of (\ref{alpha}), which Mathematica is able to 
do easily. Notice that here the recursion 
relation is between coefficients in different dimensions but of the same order (coefficients of 
$T^d$ vs. those of $T^{d+2}$ but not of $T^{2d}$ which are of next order except for $d=2$).
Unfortunately we did not succeed in computing the next order term $T^{2d}$ using the same idea.

\section{Second correction: $(T/k)^{2d}$}

Going to the second order correction, taking into account \eqref{eq:wijcond} and \eqref{hinf}, 
we define

\beq
\label{eq:w2ijdef}
w_{ij}^{(2)}(q) =\left\{\begin{array}{lcc} \int^q_0 dz\,z\;\left(u^{(2)}(z)\;b_{ij}(z) 
+u^{(1)}(z)\; b_{ik}(z)\; w_{kj}^{(1)}(z)\right)\quad&,&\quad (ij)\neq (12)\\\\
-\int^\infty_q dz\,z\;\left(u^{(2)}(z)\;b_{12}(z) 
+\;u^{(1)}(z)\; b_{1k}(z)\; w_{k2}^{(1)}(z)\right)
\quad&,&\quad (ij)= (12)
\end{array}\right.
\eeq

\noi
With these definitions we assure that the conditions that define the basis $h_\pm(z)$,

\bea
\label{eq:w2ij=0}
w^{(2)}_{11}(0) &=& w^{(2)}_{21}(0) = w^{(2)}_{22}(0) = 0\quad,\cr
w^{(2)}_{12}(\infty) &=& 0
\eea

\noi 
hold. To get the correction to the propagator, we focus on $w_{12}^{(2)}(q)$ and its limit 

\beq
\label{eq:w212q=0}
w_{12}^{(2)}(q)\quad \stackrel{q\rightarrow 0}{\longrightarrow}\quad 
 G_2^{(0)}(k)\, g_2(\omega/k) + a\,q^{-2\,\nu + 2\,d}\,\left(1+{\cal O}(q)\right)\quad.
\eeq  

\noi
For $\nu<d$ the correction to the propagator is

\beq
\label{eq:g2}
g_2(\omega/k) =A_2\alpha_{2d}+B_2\alpha_{2d+2} + 
A_1{}^2\,g_{dd}
+ B_1{}^2\,g_{d+2,d+2}
+2 A_1\,B_1\,g_{d,d+2}\quad,
\eeq

\noi
with

\beq
\label{eq:I1}
g_{rs} = \frac{1}{2}\int_0^\infty dz\,\left(F_{rs}(z)+F_{sr}(z)\right)\quad,
\eeq

\noi
whose integrand is explicitly given by

\beq
\label{Frs}
\frac{\pi}{2\sin{(\pi\nu)}}F_{rs}(z)= 
z^{r-1}\,I_\nu(z)\,K_\nu(z)\, I_s[K_\nu^2;z,\infty]-
z^{r-1}K^2_\nu(z)\,I_s[I_\nu K_\nu;z,0]\quad.
\eeq


The behaviour of the integral can be read from the behaviour of the Bessel functions, 
which main properties are summarised in appendix \ref{app:someuseful}. 
While for $z\to\infty$ both terms in (\ref{Frs}) are suppressed by $e^{-2\,z}$, for $z\to0$ 
we have

\beq
F_{rs}(z)\sim z^{-2\,\nu +r+s-1}\quad.
\eeq

So, the integral \eqref{eq:I1} converges if $2\,\nu<r+s$. 
Since we need $r+s=2\,d,2\,d+2,2\,d+4$, for $\nu<d$ no substractions are needed, and the 
contribution to the propagator at order $(T/k)^{2d}$, i.e. $g_2(\omega/k)$, can be evaluated directly. 

This procedure can be continued for larger $\nu$ using proper subtractions \eqref{gT}. For example, 
for $\nu<d+1$ one gets the same formula \eqref{eq:g2} but now with 

\beq
\label{Idd0minus}
g_{dd}=\int_0^\infty dz\left(F_{dd}(z)-\frac{2^{2\nu}}{2\nu d(d-2\nu)}
\frac{\Gamma(\nu)}{\Gamma(-\nu)}z^{-2\nu+2d-1}\right)\quad,
\eeq


\noi
with all the rest the same.

More generally, for $2\nu>r+s$, the renormalised formula \eqref{eq:I1} looks like

\begin{align}
\label{popravek}
&g_{rs}=\frac{1}{2}\int_0^\infty dz\left[F_{rs}(z)
-\sum_{n=0}^{\floor{2\nu-r-s}}z^{-2\nu+r+s-1+n}
\frac{\sqrt{\pi } \csc (\pi 
   \nu )}{4 \Gamma (n+1)
   \Gamma (n-\nu +1)}\times\right.\non\\
&\left(\frac{\pi 
   4^{\nu } 
   \Gamma (\nu +1)(-1)^n\,
   _5F_4\left(-n,\frac
   {1}{2}-\nu ,-n-\nu
   ,\frac{s}{2}-\nu ,\nu
   -n;\frac{1}{2}-n,1-2 \nu
   ,1-\nu ,\frac{s}{2}-\nu
   +1;1\right)}{\Gamma (n+\nu
   +1)\Gamma(\frac{1}{2}-n)(-2 \nu)
   \Gamma(1-\nu)(\frac{s}{2}-\nu)}\right.\non\\
& \left.\left.+\frac{\Gamma
   \left(n-\nu
   +\frac{1}{2}\right) \,
   _5F_4\left(\frac{1}{2},-n,
   \frac{s}{2},\nu -n,2 \nu
   -n;\frac{s}{2}+1,1-\nu ,\nu
   +1,-n+\nu
   +\frac{1}{2};1\right)}{\nu 
   s \Gamma (n-2 \nu
   +1)}\right)\right]\non\\
&+\left(r\leftrightarrow s\right)\quad.
\end{align}

\noi
Notice that both $_5F_4$ are actually finite sums.

\subsection{\label{nuvminusnu}$\nu\to-\nu$}

A check of the results is the behaviour of the propagator for $\nu\to-\nu$ (or, 
equivalently, $\Delta\to d-\Delta$). In this case one should get the inverse propagator:

\beq
\label{G2minusnu}
G_2(-\nu)=\frac{1}{G_2(\nu)}\quad.
\eeq

\noi
Since at this order 

\beq
G_2(\nu)=G_2^{(0)}(\nu)\left(1+\frac{g_1(\nu)}{(kz_h)^d}+\frac{g_2(\nu)}{(kz_h)^{2d}}\right)
\eeq

\noi
and since due to (\ref{eq:propT=0}) indeed 

\beq
G_2^{(0)}(-\nu)=\frac{1}{G_2^{(0)}(\nu)}\quad,
\eeq

\noi
eq. (\ref{G2minusnu}) implies

\beq
g_1(-\nu)=-g_1(\nu)\quad,
\eeq

\noi
which is easily satisfied by (\ref{eq:w1012}), and 

\beq
\label{g2minusnu}
g_2(-\nu)=g_1^2(\nu)-g_2(\nu)\quad.
\eeq

\noi
To check this one explicitly, let us first rewrite (\ref{eq:I1}) in terms of the Wronskian 

\beq
\label{grs}
g_{rs}=\frac{1}{2}\alpha_r\alpha_s- \frac{1}{4\nu}\int_0^\infty dz\,\left(W(F_r,G_s;z)+W(F_s,G_r;z)\right)
\eeq

\noi
and (see definitions in appendix \ref{app:someuseful})

\bea
F_r(z)&=&\frac{z^r}{r}\left(2f_r(z)+g_r^{(+)}(z)+g_r^{(-)}(z)\right)+2\nu\alpha_r\quad,\\
G_r(z)&=&\frac{z^r}{4\nu r}\left(g_r^{(+)}(z)-g_r^{(-)}(z)\right)\quad,
\eea

\bea
F_r'(z)&=&-\frac{4\nu}{\pi}\sin{(\pi\nu)}z^{r-1}K_\nu^2(z)\quad,\\
G_r'(z)&=&\frac{z^{r-1}}{2}K_\nu(z)\left(I_\nu(z)+I_{-\nu}(z)\right)\quad.
\eea

\noi
From (\ref{alpha}), (\ref{eq:fgpm}) and the above definitions it is straightforward to see that 
under $\nu\to-\nu$ the quantity $\alpha_r$ is odd, while $A_m$, $B_m$, $F_r$ and $G_r$ are even. 
Then in (\ref{g2minusnu}) the $A_2$, $B_2$ as well as the integrals of Wronskians exactly drop out, 
while the $\alpha_r\alpha_s/2$ terms in (\ref{grs}) take care of the 
$g_1^2$ term in (\ref{g2minusnu}). The relation (\ref{G2minusnu}) is 
indeed correct up to order $T^{2d}$.

\section{Special $d$}

So far the only check we did is the analytic comparison with the large $\nu$ solution 
of \cite{Rodriguez-Gomez:2021pfh}. Now we will specialise to the $d=2$ case which is 
exactly known. We will compare our numerical solution with this exactly known result and 
find agreement at second order $T^4$. Then we will pass on to $d=4$ and predict the 
second order $T^8$ correction to the propagator.

\subsection{$d=2$}

Equation (5.8) of \cite{Grozdanov:2019uhi} in our euclidean setting  is

\beq
\label{eq:Sashoprop}
G(k) = C_\Delta
 \frac{\Gamma(z_+ +\frac{\Delta}{2})}{\Gamma(z_+ +1-\frac{\Delta}{2})}\,
\frac{\Gamma(z_- +\frac{\Delta}{2})}{\Gamma(z_- +1-\frac{\Delta}{2})}\quad,
\eeq

\noi
where 

\beq
C_\Delta=\frac{\Gamma(-\nu)}{\Gamma(\nu)2^{2\Delta-2}}\left(\frac{2}{z_h}\right)^{2\Delta-2}
\eeq

\noi
is a normalisation constant, and 

\beq
z_\pm\equiv \frac{z_h}{2}\,\left(\omega\pm i\, q\right)\qquad,\qquad
z_+\,z_- = \frac{( kz_h)^2}{4}\quad.
\eeq

\noi
We need the expansion for large $ kz_h$, i.e. large $|z_\pm|$. 
To get it, we use results in \cite{Tricomi:1951}. We have for large $|z|$
\beq\label{eq:TricomiErdely}
\frac{\Gamma(z+a)}{\Gamma(z+b)}\sim 
z^{a-b}\,\sum_{n\in\mathbb N} c_n(a,b)\,\frac{z^{-n}}{n!}\qquad {\rm with}\qquad 
c_n(a,b)\equiv(-)^n\,(b-a)_n\,B_n^{(1+a-b)}(a) \quad,
\eeq

\noi
where $B_n^{(k)}(x)$ are generalized Bernoulli polynomials \cite{Tricomi:1951}.

Applying to \eqref{eq:Sashoprop}, taking into account that 
\beq
c_{2n}=(1-\Delta)_{2n}\,B_{2n}^{(\Delta)}\left(\Delta/2\right)\qquad,\qquad
c_{2n+1}=-(1-\Delta)_{2n+1}\,B_{2n+1}^{(\Delta)}\left(\Delta/2\right)=0\quad,
\eeq
we get the expansion
\bea\label{eq:Gsexp}
G(k)&\sim& C_\Delta\,(z_+\,z_-)^{\Delta-1}\,
\left(1 + \frac{c_2}{2!\,z_+{}^2} + \frac{c_4}{4!\,z_+{}^4}+ {\cal O}(T^6)\right)\,
\left(1 + \frac{c_2}{2!\,z_-{}^2} + \frac{c_4}{4!\,z_-{}^4}+{\cal O}(T^6)\right)\cr
&=&G^{(0)}(k)\,\left(1 + \frac{c_2}{2\,z_+{}^2} + \frac{c_2}{2\,z_-{}^2} 
+ \frac{c_4}{24\,z_+{}^4} + \frac{c_2{}^2}{4\,z_+{}^2\,z_-{}^2}+ \frac{c_4}{24\,z_-{}^4}+ {\cal O}(T^6)\right)\quad,
\eea

\noi
where

\beq
c_2 = -\frac{1}{12}\,\prod_{l=0}^2(\Delta - l)\qquad,\qquad
c_4 = \frac{1}{240}\,(5\,\Delta+2)\,\prod_{l=0}^4(\Delta - l)\quad.
\eeq

From \eqref{eq:Gsexp} we find that the $T^2$-correction to the propagator

\beq
\frac{g_1(\omega/k)}{(kz_h)^2}= \left(\frac{2\pi T}{ k}\right)^2\,\frac{1}{3}\,\Delta(\Delta-1)(\Delta-2)\,
(1-2\,(\omega/k)^2)
\eeq

\noi
coincides with our result \eqref{eq:w1012} at $d=2$, 
while the $T^4$-correction is

\bea
\label{eq:sashoT^4}
\frac{g_2(\omega/k)}{(kz_h)^4}&=&
\left(\frac{2\pi T}{ k}\right)^4\,\frac{1}{90}\,\Delta(\Delta-1)(\Delta-2)\\
&\times&\left(
9\,\Delta\,(\Delta-2) -12 + (5\,\Delta+2)\,(\Delta-3)\,(\Delta-4)\,(1-2\,(\omega/k)^2)^2
\right)\quad.\non
\eea

Writing 

\beq
\label{g2gamma}
g_2(\omega/k)=\gamma_0+\gamma_2(\omega/k)^2+\gamma_4(\omega/k)^4
\eeq

\noi
and comparing $\gamma_i$ from the numerical evalulation of the integrals \eqref{eq:g2} 
(with \eqref{Idd0minus} instead of \eqref{eq:I1} for $2<\nu<5/2$) as functions of 
$\nu\in\left[1,5/2\right]$ with the known analytic form \eqref{eq:sashoT^4} 
we get good agreement, see figs. \ref{g024}.

\begin{figure}[htb]
\begin{center}
\includegraphics[width=5.cm]{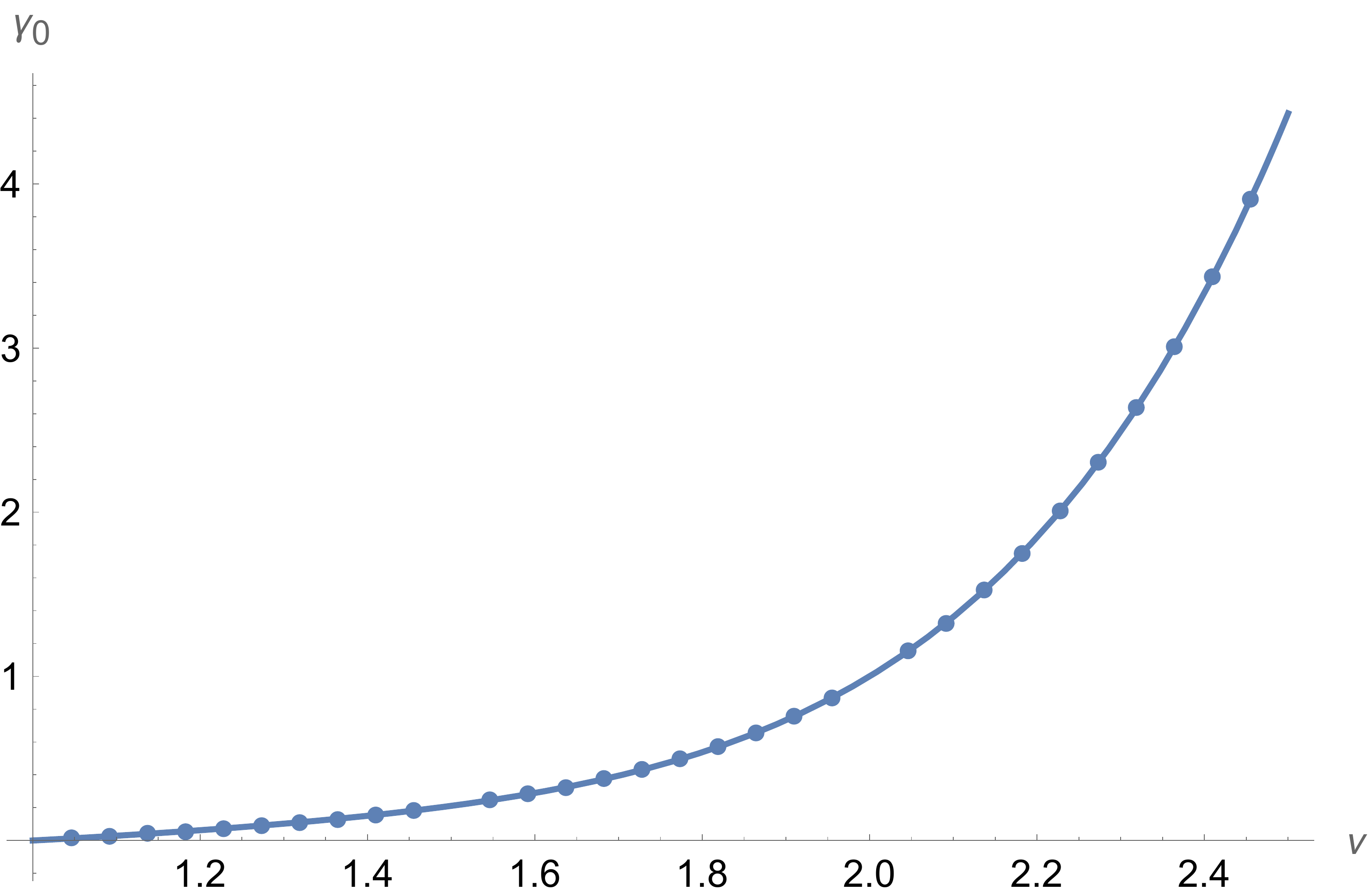}
\includegraphics[width=5.cm]{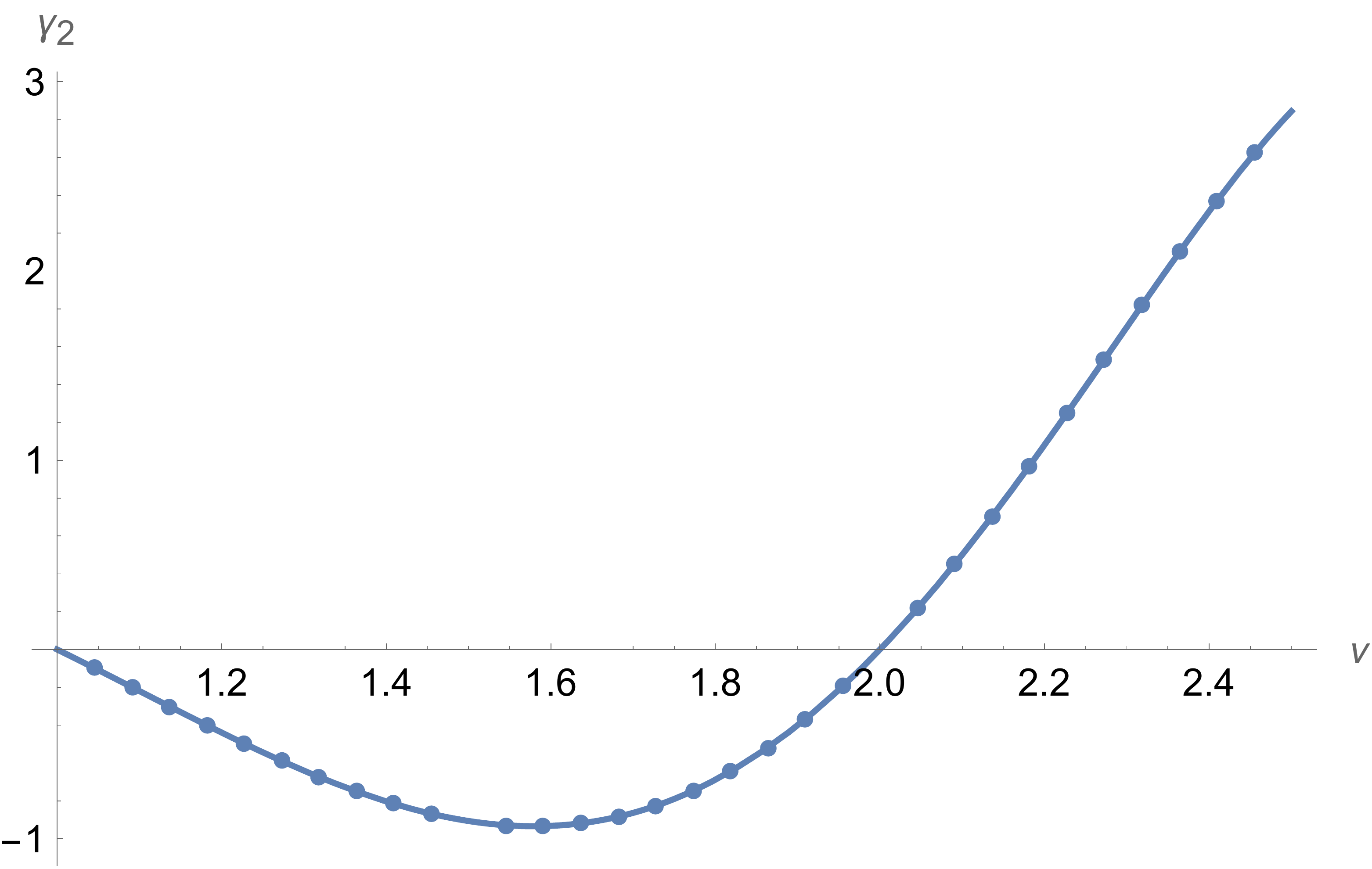}
\includegraphics[width=5.cm]{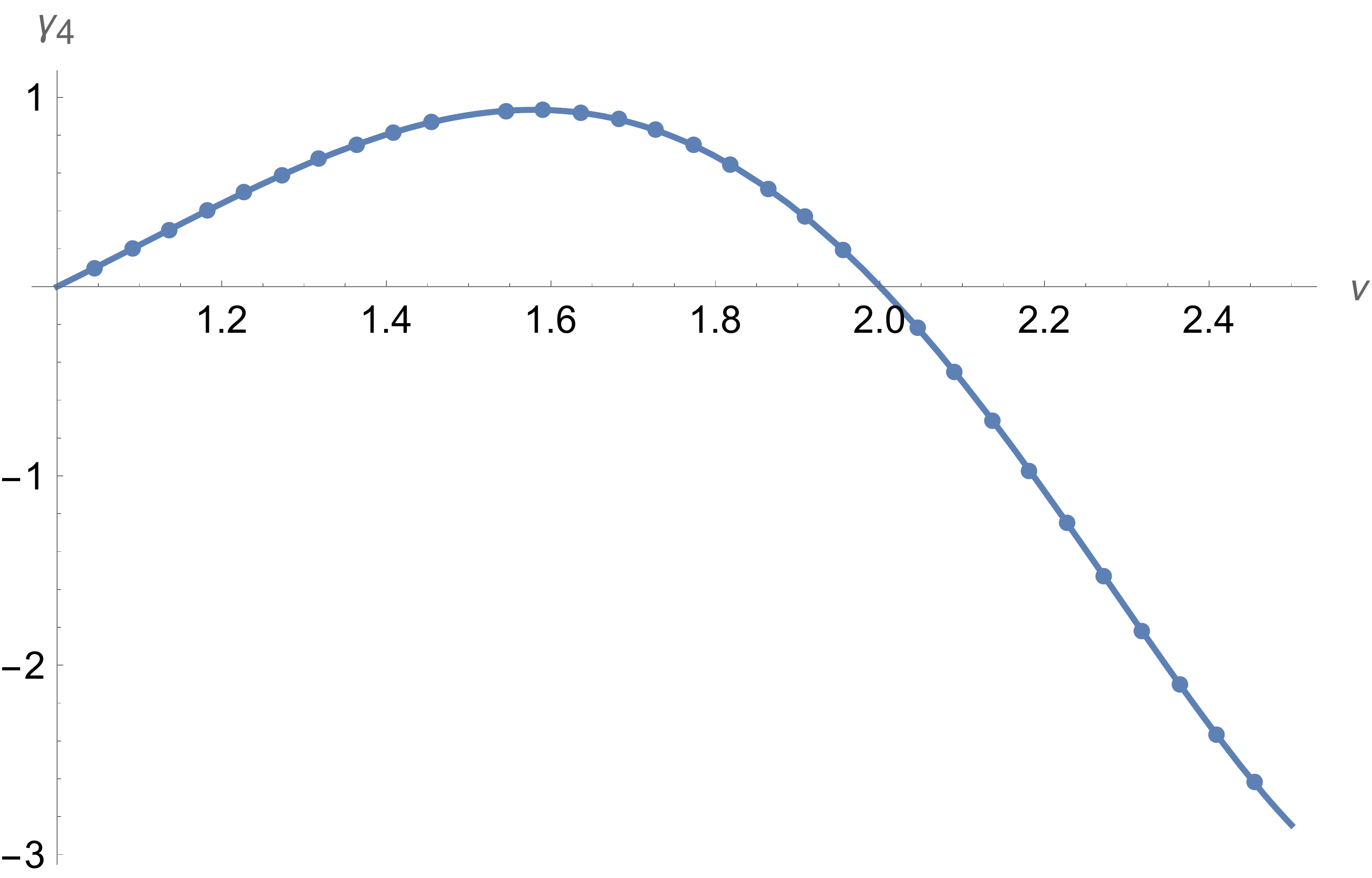}
\caption{\label{g024} Comparison of $\gamma_i$ defined in \eqref{g2gamma} between the exact result 
\eqref{eq:sashoT^4} (continuous curves) and our numerical evaluation of. eq. \eqref{eq:g2} (discrete points) 
as functions of $\nu\in\left[d/2,d+1/2\right]$ for the order $T^{2d}$ correction to the Euclidean 
propagator in the $d=2$ case.
}
\end{center}
\end{figure}

Since the expansion eq. (4.5) in \cite{Policastro:2001yb} for the $D1/D5$-branes system agrees with the full result 
\cite{Grozdanov:2019uhi} by identifying $\Delta=1+\nu$ and $q=0$, we obviously agree with \cite{Policastro:2001yb} too.

\subsection{$d=4$}

Here we first notice that our $T^4$ analytic result  (\ref{eq:sigmad=4}) completely agrees with the first nontrivial term in 
\cite{Policastro:2001yb} by identifying $\Delta=2+\nu=l+4$ and $\vec{q}=\vec{0}$. 
Next we compare our $T^8$ numerical result (\ref{eq:g2}) for $g(1)$ with eq. (2.3) of \cite{Policastro:2001yb} with the same identification as above, 
finding a perfect numerical agreement as shown in Fig. \ref{WEvsPS}.

\begin{figure}[htb]
\begin{center}
\includegraphics[width=15.cm]{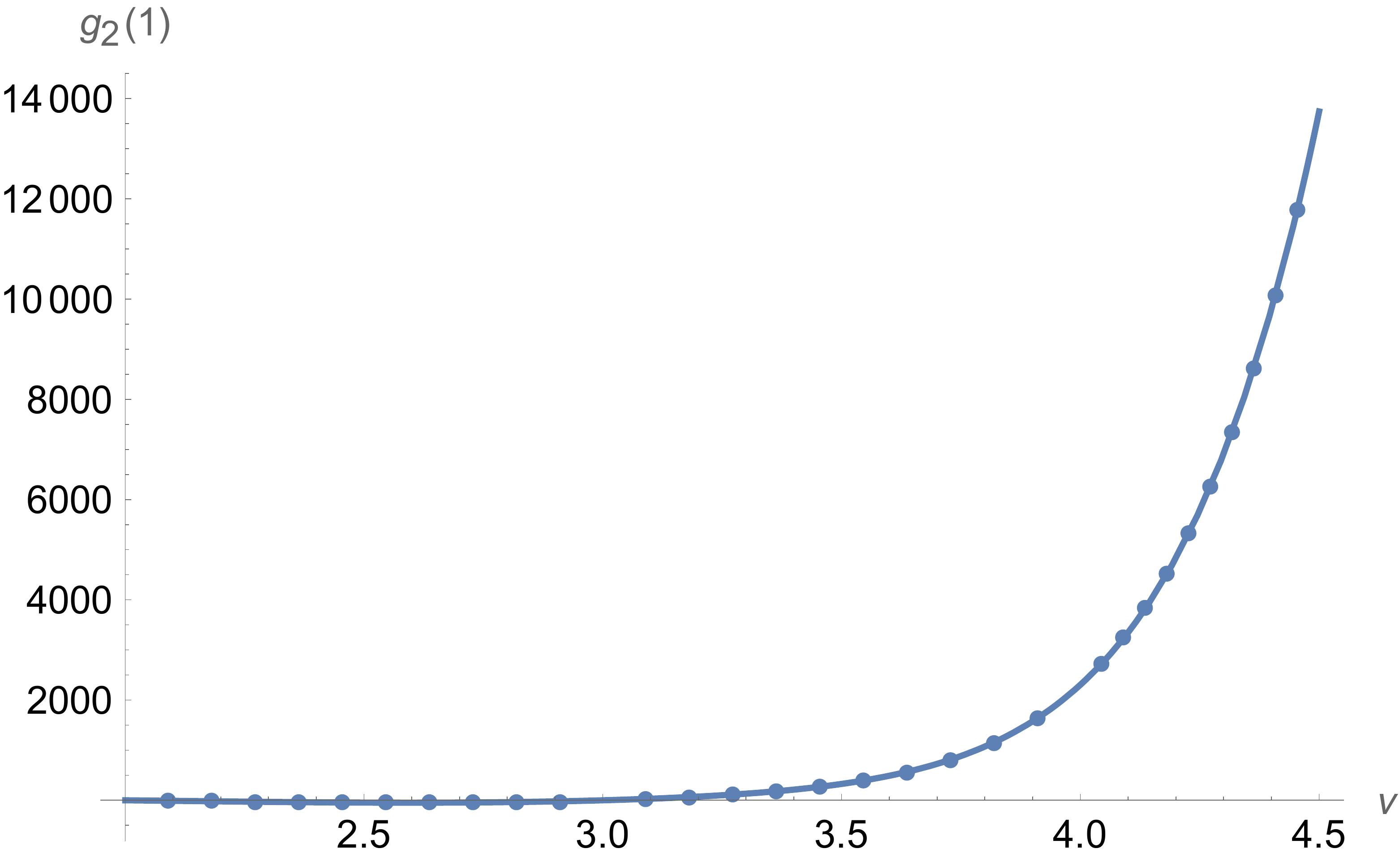}
\caption{\label{WEvsPS} The function $g_2$ at $k=\omega$ for the $d=4$ Euclidean 
propagator at order $T^8$ in our numerical evaluation of eq. \eqref{eq:g2} (discrete points) as 
functions of $\nu$ compared to the eq. (2.3) of \cite{Policastro:2001yb}  (full line).}
\end{center}
\end{figure}

Reinvigorated by this matching at $d=4$ with the case $\omega/k=1$ and $\nu=l+2$ of  \cite{Policastro:2001yb}, 
and the matching in the previous section between the exactly known $d=2$ result and our 
numerical evaluation of the second order contribution to the Euclidean propagator as a function of  
$\nu$, which is summarised in fig. \ref{g024}, we go now to the arbitrary $\omega/k$ and $d=4$ case. For the same 
$\gamma_i$, $i=0,2,4$ functions of $\nu$ we find the $T^8$ contributions in fig. \ref{gg024}.

Here the full $\nu$ dependent result is not known, but for large $\nu$ the approximate values can be 
found in \cite{Fitzpatrick:2019zqz,Rodriguez-Gomez:2021pfh} and is here taken from (\ref{eq:orden2bislargenu})\footnote{
Here there is an ambiguity of how to write the 
leading term, i.e. $\propto\nu^{10}$ or $\propto\Delta^{10}$: they are equivalent at large $\nu$ but 
make a difference for small $\nu$. We choose a conservative approach and wrote this contribution as 
a function of $\nu$, which is closer to our numerical result.}:

\beq
\label{g2largenu}
g_2^{{\rm large}\;\nu}(\omega/k)=\frac{2}{225}\nu^{10}\left(4\frac{\omega^2}{k^2}-1\right)^2\quad.
\eeq

It is interesting to know how far is the approximate result (\ref{g2largenu}) from our numerical evaluation. 
This is compared in fig. \ref{gg024nu}, which shows that indeed the large $\nu$ approximation 
is not suitable for small $\nu$, which makes our computation all the more important. On the other side, 
it should be stressed here that our method is not suitable for large $\nu$ calculations: in fact we need to 
explicitly subtract the divergent pieces and, as we have seen, their number increases linearly with 
$\nu$. 
%
%
So the two methods, the geodesic approximation 
used in \cite{Rodriguez-Gomez:2021pfh,Rodriguez-Gomez:2021mkk} for large $\nu$ and our 
method here, should be considered as complementary rather than competitive.

Said that, we tried to analytically take the limit $\nu\to\infty$. Although we were unable to prove it 
definitely, we have some indications to conjecture that the large $\nu$ result for an arbitrary dimension 
$d$ is

\bea
\label{g2conjecture}
g_2^{{\rm large}\;\nu}(\omega/k)&=&
\lim_{\nu\to\infty}\frac{1}{2}\left(A_1\alpha_d+B_1\alpha_{d+2}\right)^2\non\\
&=&\frac{2^{2d-5}\Gamma^4(d/2)}{(d+1)^2\Gamma^2(d)}\nu^{2d+2}
\left(d\left(\frac{\omega}{k}\right)^2-1\right)^2\quad.
\eea

This result coincides with (\ref{g2largenu}) for $d=4$ and with the large $\nu$ limit of 
(\ref{eq:sashoT^4}) for $d=2$. The reason why the extreme large limit is maybe doable is that 
one does not need to do it numerically, and analytical approximation of Bessel functions 
at large $\nu$ can be used \cite{NIST:DLMF}.

Notice that (\ref{g2conjecture}) confirms the exponentiation property of the propagator at large $\Delta$ 
\cite{Rodriguez-Gomez:2021pfh}.

\begin{figure}[htb]
\begin{center}
\includegraphics[width=5.cm]{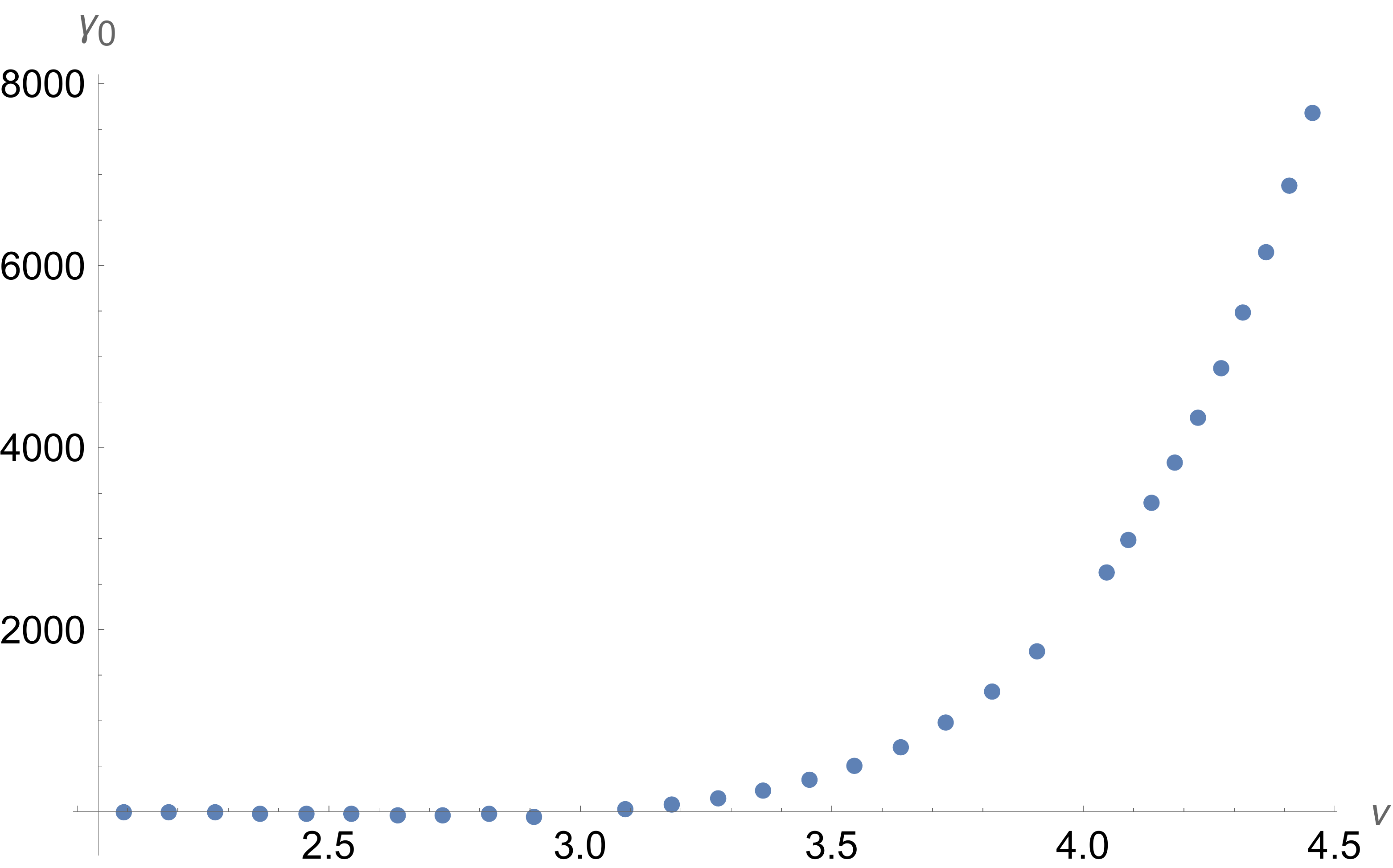}
\includegraphics[width=5.cm]{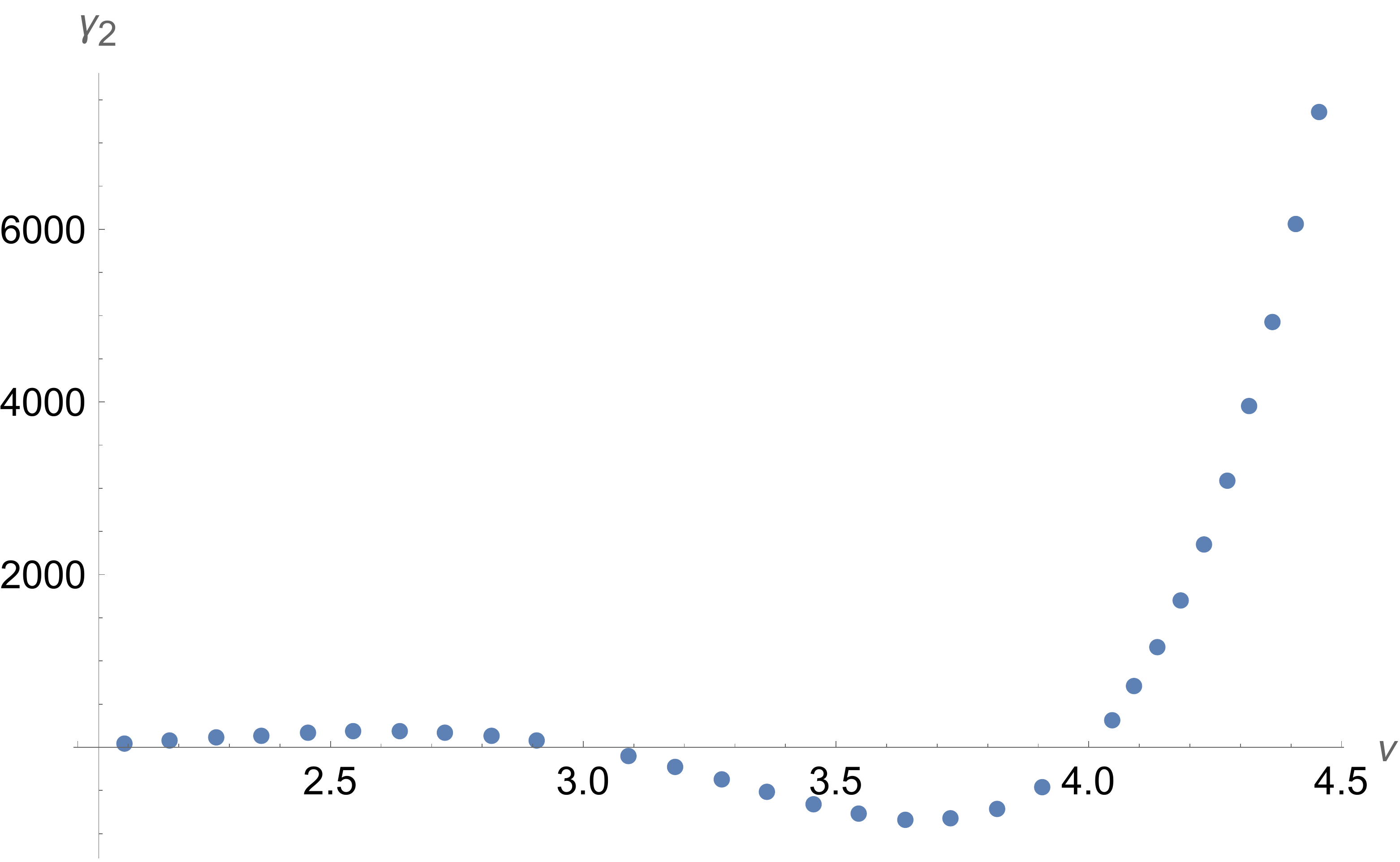}
\includegraphics[width=5.cm]{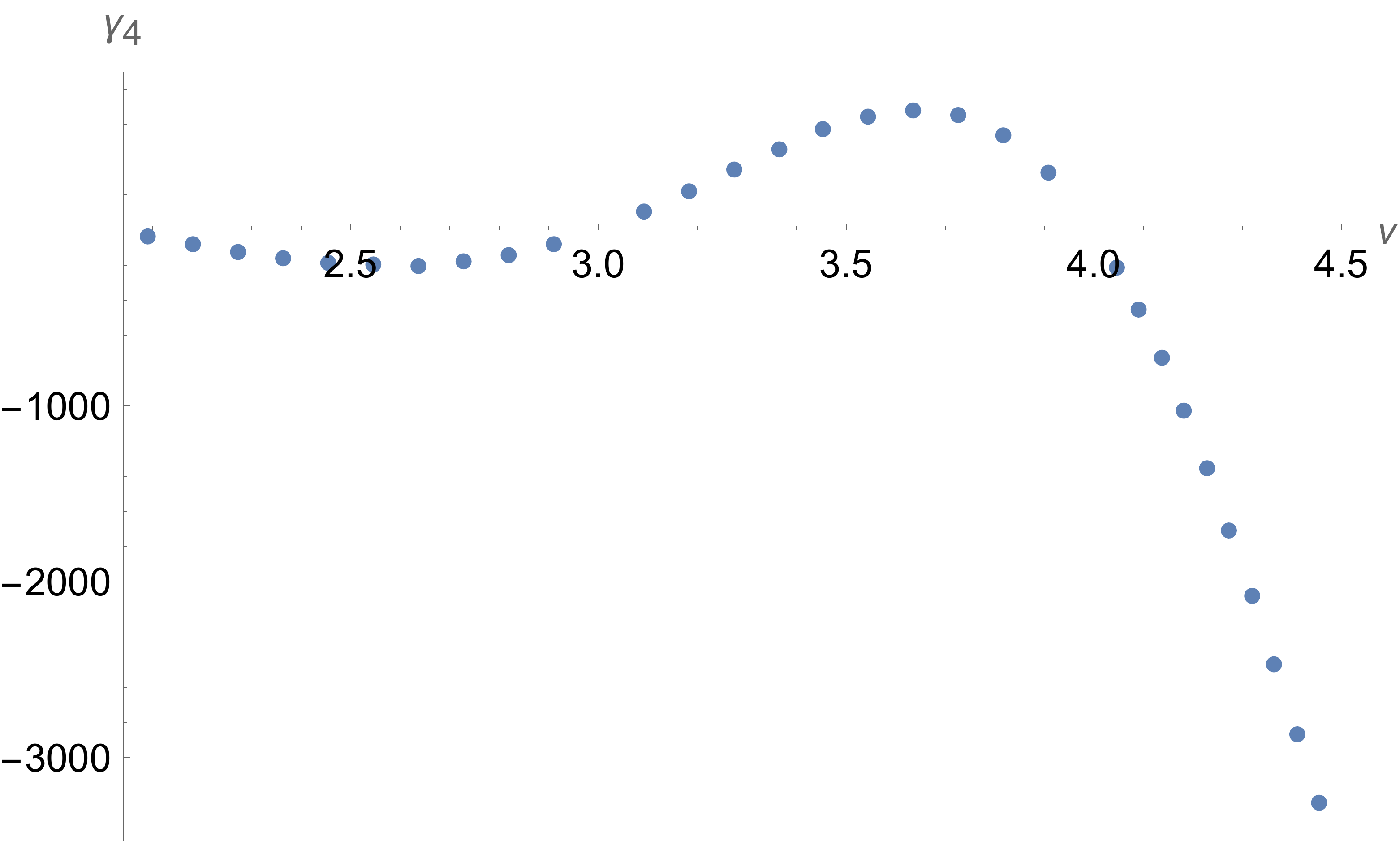}
\caption{\label{gg024} The functions $\gamma_i$ defined in \eqref{g2gamma} for the $d=4$ Euclidean 
propagator at order $T^8$ in our numerical evaluation of eq. \eqref{eq:g2} (discrete points) as 
functions of $\nu\in\left[d/2,d+1/2\right]$.
}
\end{center}
\end{figure}

\begin{figure}[htb]
\begin{center}
\includegraphics[width=5.cm]{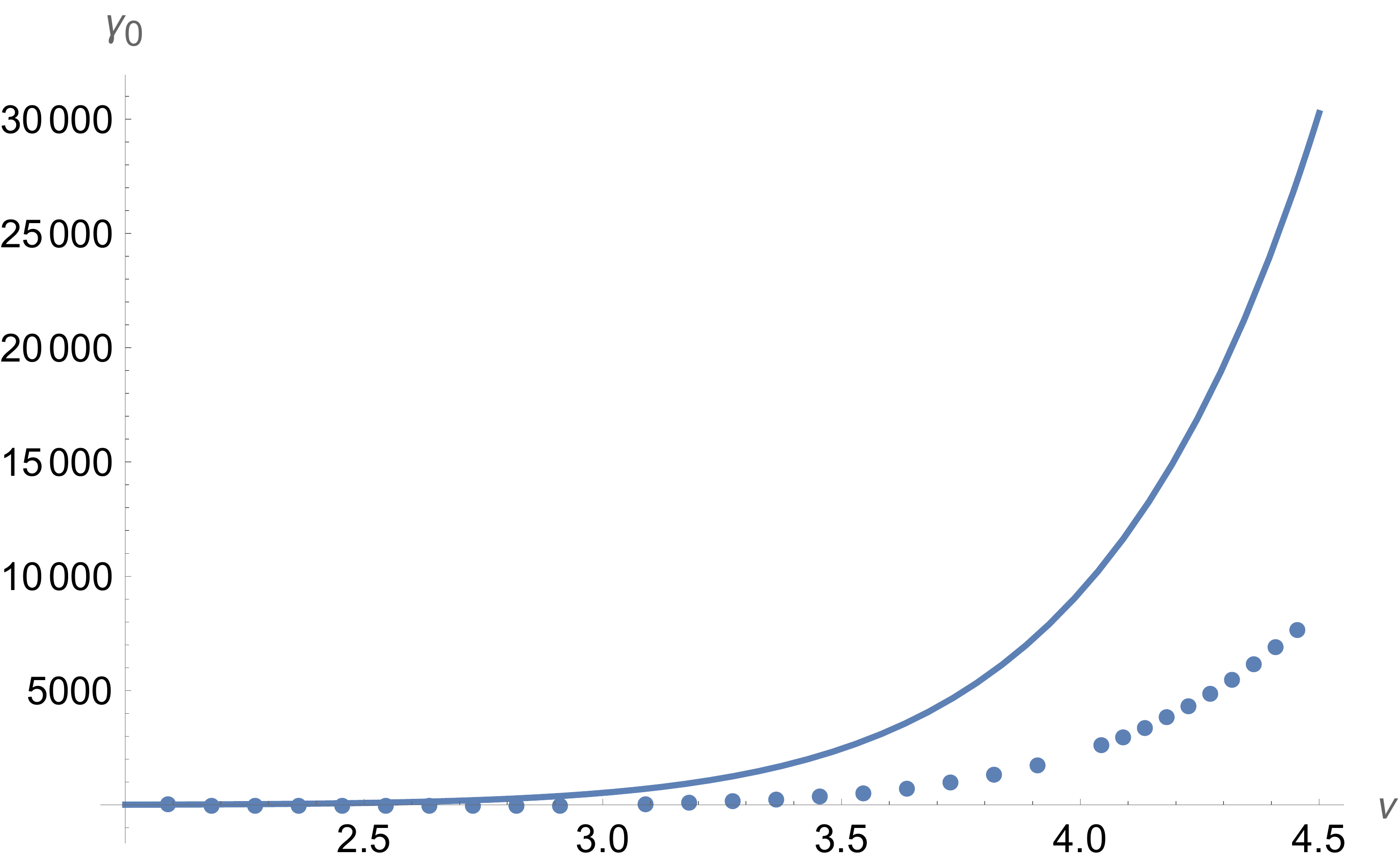}
\includegraphics[width=5.cm]{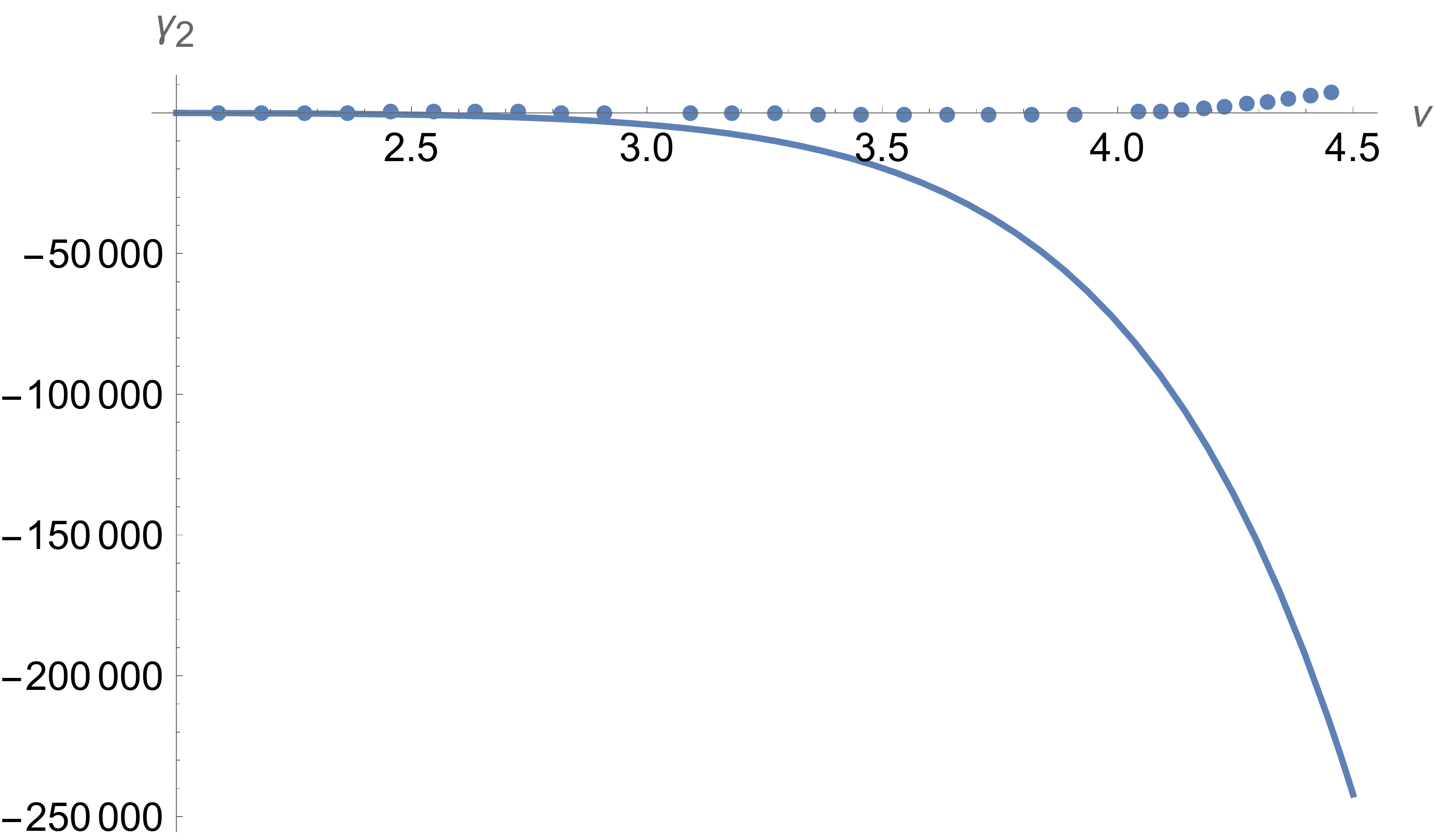}
\includegraphics[width=5.cm]{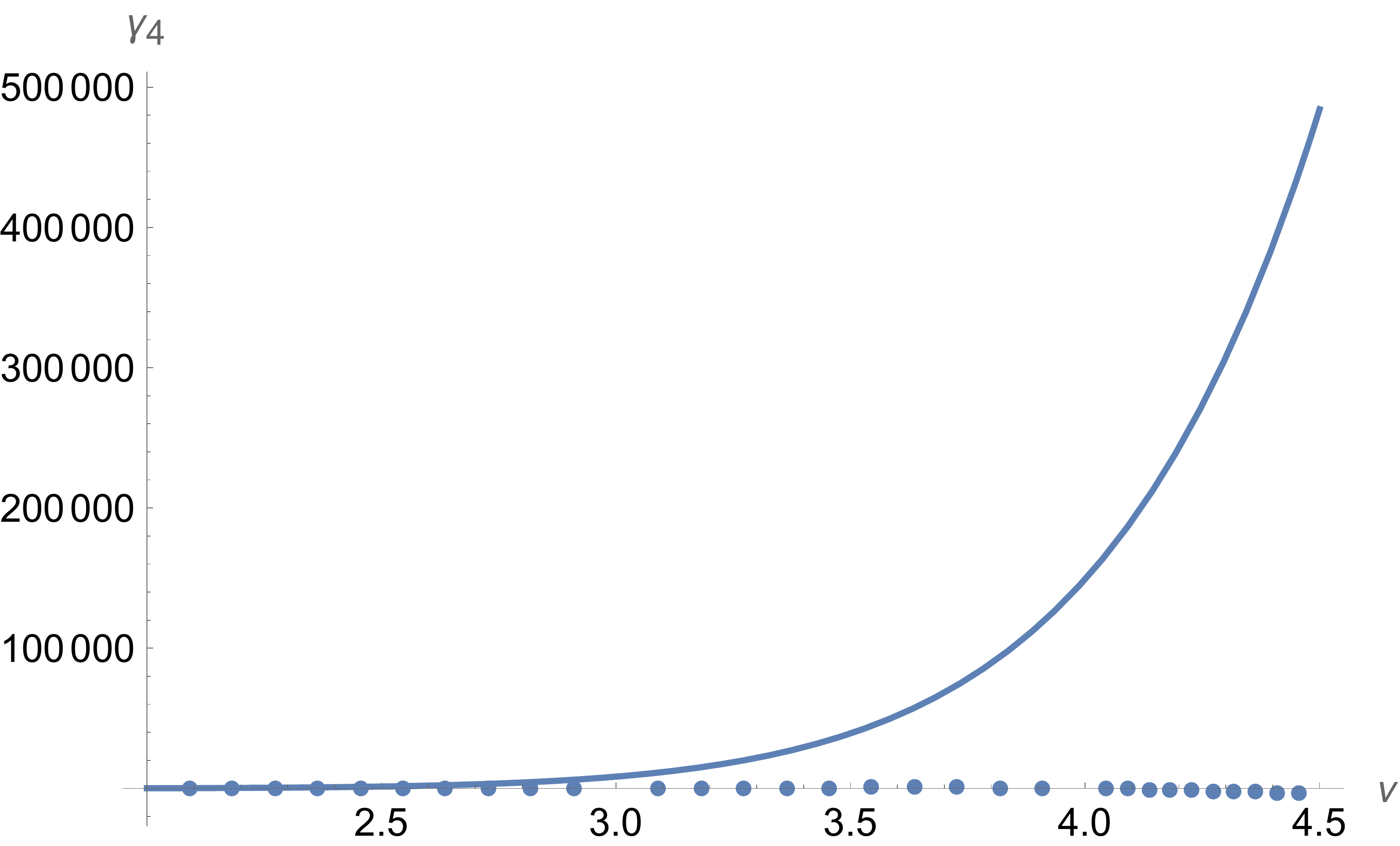}
\caption{\label{gg024nu} Comparison of $\gamma_i$ defined in \eqref{g2gamma} for the $d=4$ 
Euclidean propagator at order 
$T^8$ between the approximate result valid for large $\nu$ of \cite{Rodriguez-Gomez:2021pfh} and our 
numerical evaluation of eq. \eqref{eq:g2} (discrete points) as functions of $\nu\in\left[d/2,d+1/2\right]$.
}
\end{center}
\end{figure}

\section{Conclusions}

We showed in this paper the way to calculate the euclidean thermal propagator in a CFT using 
the AdS/CFT correspondence. Although the full formula for it has been recently found \cite{Dodelson:2022yvn}, 
it is not suited for a small temperature expansion. 
We found the first two corrections, the first analytical of order $T^d$ in (\ref{eq:w1012}), 
and the second numerical of order $T^{2d}$ via (\ref{eq:g2}) and (\ref{popravek}). The order $T^d$ 
agrees with \cite{Parisini:2022wkb} for general $d$ and with \cite{Fitzpatrick:2019zqz,Rodriguez-Gomez:2021pfh} for $d=4$ 
(summarised in (\ref{eq:sigmad=4})). The order $T^{2d}$ reproduces the well  known 
result \cite{Grozdanov:2019uhi} for $d=2$. 
This last check is summarised in fig. \ref{g024}. We then predicted the $T^8$ contributions in the 
$d=4$ case in fig. \ref{gg024}. Finally we showed in fig. \ref{gg024nu} explicitly that the 
large $\nu$ approximation does not properly describe the small $\nu$ limit, a result which is 
not surprising but had to be checked. The method presented here is thus appropriate 
for this range of $\nu$, not described by the geodesic approximation.

Our approach can deal with general spacetime dimensions $d$. It is not tied to the 
Heun equation as is the case of the analytical solution \cite{Dodelson:2022yvn}. 
In fact in $d>4$ apparently more singular points are present and so it is not clear if the equation 
to be solved is still of the Heun type. In this paper we found for arbitrary $d$ an explicit expression 
(\ref{eq:w1012}) for the propagator at order $T^d$ for arbitrary $\nu$. At the order $T^{2d}$ we conjecture 
an analytic formula (\ref{g2conjecture}), valid for large $\nu$. 

The perturbative expansion in $T/k$, which we do in 
$d$-dimensional Euclidean spacetime, seems to be compatible with the analytical continuation 
to Minkowski spacetime. This is checked for example if we compare our Euclidean result with the 
Minkowski all-order result in $d=2$.

Few words of caution are perhaps important at this point. As studied in \cite{Policastro:2001yb}, 
the Heun equation has one potential problem and one real problem. The potential one is connected to the 
uniform convergence of the asymptotic expansion in the whole coordinate interval, also close to the singularities. 
The authors of \cite{Policastro:2001yb} use successfully the method of Langer and Olver \cite{Langer,Olver}. 
In our case this would roughly mean that the coefficients (\ref{wij})-(\ref{w120}) need to be finite, which we easily prove they are. 
More precisely, in the limit $k=\omega$ we start from the same equation of \cite{Policastro:2001yb}. The difference with them is that we 
automatically find an expansion in positive integer powers of $T^{d}$, while in \cite{Policastro:2001yb} this turns out to be true 
only after some cancellations. For our more general case of $k\ne\omega$ we do not know of any method equivalent to Langer and Olver. 
Finding such a method is clearly beyond the scope of this paper.  We however check our results by comparing them to the known 
ones in the literature and find perfect agreement whenever available. 
The second, real problem, is the appearance in the solution of terms 
proportional to positive powers of $\exp{(-1/T)}$. These terms, as in \cite{Policastro:2001yb}, 
are not available in our method. We would like here to stress that these terms are certainly one of the reasons why the 
full solution \cite{Dodelson:2022yvn} mentioned above is not suited for a low $T$ expansion.


\subsubsection*{Acknowledgments}
We thank Sa\v so Grozdanov for many illuminating discussions on subjects relating this 
work, Vasil Avramov for pointing out and clarifying some aspects of reference \cite{CarneirodaCunha:2015qln}, 
Jorge Russo for very important remarks and Enrico Parisini, Kostas Skenderis, Benjamin Withers and 
Kuo-Wei Huang for useful correspondence. 
BB acknowledges the financial support from the Slovenian 
Research Agency (research core funding No.~P1-0035). 
AL acknowledges the financial support from CONICET through PIP 02229 and PUE084, and from La Plata University-11/X910.
BB (AL) thanks the Physics Department 
of La Plata University (Jo\v zef Stefan Institute) for hospitality.

\appendix

\section{Some useful identities}
\label{app:someuseful}

We first present some useful formulae regarding Bessel functions for indices 
$\nu\notin \mathbb Z$.
By using the expansions given in \cite{gradshteyn2007}, we have for $q\ll 1$, 
\bea\label{eq:IKq<<1}
I_\nu(q)&=&
= \frac{1}{\Gamma(1+\nu)}\,\left(\frac{q}{2}\right)^\nu\;
_0{}F_1\left(1+\nu;\frac{q^2}{4}\right)\quad,
\cr\cr
&&\cr
K_\nu(q) &=& \frac{\Gamma(\nu)}{2}
\left(\frac{q}{2}\right)^{-\nu}\;
_0{}F_1\left(1-\nu;\frac{q^2}{4}\right) + \frac{\Gamma(-\nu)}{2}
\;
\left(\frac{q}{2}\right)^\nu\;
_0{}F_1\left(1+\nu;\frac{q^2}{4}\right)\quad,\cr
&&
\eea
and for  $q\gg 1$ the asymptotic expansions is
\bea\label{eq:IKq>>1}
I_\nu(q)&\sim&\frac{e^q}{\sqrt{2\,\pi\,q}}\;
_2{}F_0\left(\frac{1}{2}+\nu,\frac{1}{2}-\nu;\frac{1}{2\,q}
\right) 
+ i\,e^{ i\,\pi\,\nu}\frac{e^{-q}}{\sqrt{2\,\pi\,q}}\;
_2{}F_0\left(\frac{1}{2}+\nu,\frac{1}{2}-\nu;-\frac{1}{2\,q}
\right)\quad,\cr\cr
&&\cr
K_\nu(q) &\sim& \sqrt{\frac{\pi}{2\,q}}\;e^{-q}\;
_2{}F_0\left(\frac{1}{2}+\nu,\frac{1}{2}-\nu;-\frac{1}{2\,q}
\right)\quad.
\eea

The elements of the matrix ${\bf b}(q)$ in \eqref{eq:b} are defined by,
\bea
\label{XY}
2\,\nu\,b_{11}(q)&\equiv& 2\,\nu\,I_\nu(q)\,K_\nu(q)\cr
&=& 
{}_1F_2\left({\frac{1}{2}\atop 1+\nu,1-\nu};q^2\right)
+ \frac{\Gamma(-\nu)}{\Gamma(\nu)}\,
\left(\frac{q}{2}\right)^{2\,\nu}\,
{}_1F_2\left({\frac{1}{2}+\nu\atop 1+2\,\nu,1+\nu};q^2\right)\quad,\cr
&&\cr
b_{22}(q)&\equiv&-b_{11}(q)\quad,\cr
&&\cr
\frac{4}{\alpha}\,b_{12}(q) &\equiv& 4\,K_\nu(q)^2 =
\Gamma(-\nu)^2\,\left(\frac{q}{2}\right)^{2\,\nu}\,
{}_1F_2\left({\frac{1}{2}+\nu\atop 1+2\,\nu,1+\nu};q^2\right)\cr
&&\cr
&+& \Gamma(\nu)\,\Gamma(-\nu)\,
{}_1F_2\left({\frac{1}{2}\atop 1+\nu,1-\nu};q^2\right)
+ (\nu\rightarrow -\nu)\quad,\cr
-\alpha\,b_{21}(q) &\equiv& I_\nu(q)^2 = 
\frac{1}{\Gamma(1+\nu)^2}\,
\left(\frac{q}{2}\right)^{2\,\nu}\,
{}_1F_2\left({\frac{1}{2}+\nu\atop 1+2\,\nu,1+\nu};q^2\right)\quad,
\eea
where we have used \eqref{eq:IKq<<1}, the notable product identity
\beq
{}_0F_1\left(a;z\right)\;{}_0F_1\left(b;z\right)=
{}_2F_3\left({\frac{1}{2}(a+b),\frac{1}{2}(a+b-1)\atop a,b,a+b-1};4\,z
\right)\quad,
\eeq
and the fact that 
\beq
{}_{p+1}F_{q+1}\left({a_1,\dots a_p,c\atop b_1,\dots b_q, c};z\right) 
= {}_{p}F_{q}\left({a_1,\dots,a_p\atop b_1,\dots, b_q};z\right)\quad.
\eeq
%
%
%
%
%

Using 

\beq
\int^q dz\;z^{s-1}\;
{}_{p}F_{q}\left({a_1,\dots,a_p\atop b_1,\dots,b_q};z^2\right)
=
\frac{q^s}{s}\;{}_{p+1}F_{q+1}\left({a_1,\dots,a_p,\frac{s}{2}\atop b_1,\dots,b_q,1+\frac{s}{2}}; q^2\right)\quad,
\eeq

\noi
the integrals \eqref{eq:Is} are (for positive $s$ )
\bea\label{eq:intIK}
I_s[b_{11};q,0] &=& 
\frac{1}{2\,\nu}\,
\frac{q^s}{s}\left(f_s(q) + g^{(+)}_s(q)\right)\quad,\cr
I_s[b_{22};q,0] &=& -
\frac{1}{2\,\nu}\,
\frac{q^s}{s}\left(f_s(q) + g^{(+)}_s(q)\right)\quad,\cr
I_s[b_{12};q,\infty] &=& 
\,\frac{ G_2^{(0)}(k)}{2\,\nu}
\left(\frac{q^s}{s}\left(2\,f_s(q) + g^{(+)}_s(q) +g^{(-)}_s(q)\right)+2\nu\alpha_s\right)\quad,\cr
I_s[b_{21};q,0] &=& 
-\frac{1}{2\,\nu G_2^{(0)}(k)}
\frac{q^s}{s} g^{(+)}_s(q) \quad,
\eea
where we defined the functions $f_s(q)$ and $g^{(\pm)}_s(q)$ as 

\begin{align}
\label{eq:fgpm}
f_s(q)&\equiv
{}_2F_3\left({\frac{s}{2}, \frac{1}{2}\atop 1+\nu,1+\frac{s}{2},1-\nu};q^2\right)\quad,\\
g^{(\pm)}_s(q)&\equiv
\gamma_s^{(\pm)}\;q^{\pm2\,\nu}\;
{}_2{}F_3\left({\frac{s}{2}\pm\nu, \frac{1}{2}\pm\nu\atop
1\pm2\,\nu,1\pm\nu+\frac{s}{2},1\pm\nu};q^2\right)
\quad,\quad
\gamma_s^{(\pm)} \equiv \frac{2^{\mp2\,\nu}\,s}{s\pm2\,\nu}\;
\frac{\Gamma(\mp\nu)}{\Gamma(\pm\nu)}\quad,\non
\end{align}

\noi
and $\alpha_s$ in (\ref{alpha}). 
%
%
%
Eq. \eqref{eq:intIK} can be used to compute the elements of the matrix 
(\ref{eq:w1})

\beq\label{eq:Wm}
 \int_{q_0}^q dz\,z\,u^{(m)}(z)\;{\bf b}(z) = 
A_m\,I_{d\,m}[{\bf b};q,q_0] + B_m\,I_{d\,m+2}[{\bf b};q,q_0]\quad,\quad m=1,2,\dots\quad.
\eeq

\section{\label{Fourier}From $x$ to $k$ space}

In reference \cite{Rodriguez-Gomez:2021mkk},  Rodr\'iguez-G\'omez and Russo computed the low temperature expansion of the propagator up to second order in $T^4$, in four dimensions and large operator   dimension limit.  
Their result, given in coordinate space $(\tau, \vec x)$, can be written as
\beq\label{eq:Giorgioresult}
G_2(x, \eta) = G_2^{(0)}(x) + 
G_2^{(1)}(x, \eta) + G_2^{(2)}(x, \eta) +\dots\quad,
\eeq
where $\;x\equiv\sqrt{\tau^2+\vec x^2}$, $\;\eta\equiv\frac{\tau}{x}$, and
\bea\label{eq:GmGiorgio}
G_2^{(0)}(x)&=& x^{-2\,\Delta}\quad,\cr
G_2^{(1)}(x,\eta) &=& \frac{2\,\Delta}{15}\,\left(\frac{\pi\,T}{2}\right)^4\,
C_2^{(1)}(\eta)\,x^{4-2\,\Delta}\quad,\cr
G_2^{(2)}(x, \eta) &=&
\frac{2\,\Delta^2}{15^2}\,\left(\frac{\pi\,T}{2}\right)^8\,
\left(C_0^{(1)}(\eta)+C_2^{(1)}(\eta)+C_4^{(1)}(\eta)\right)\,
x^{8-2\,\Delta}\quad.
\eea
Here  $C_m^{(a)}(\eta)$ are Gegenbauer polynomials, the relevant ones are
\beq\label{eq:Gegen}
C_0^{(1)}(\eta)= 1\quad,\quad 
C_2^{(1)}(\eta)= -1 + 4\,\eta^2\quad,\quad
C_4^{(1)}(\eta)= 1 - 12\,\eta^2 + 16\,\eta^4 \quad.
\eeq

To compare \eqref{eq:Giorgioresult} with our results, we need 
\eqref{eq:GmGiorgio} in momentum space $(\omega,\vec q)$. 
To this end, we first introduce the Fourier transform
\beq
f_{2a}(k) \equiv \int d^dx\, e^{-i\,k\cdot x}\,x^{-2\,a}
= c_{2a}\,k^{2\,a-d}\quad,
\eeq
where $k\equiv\sqrt{\omega^2+\vec q^2}$. 
By using the representation
\beq
x^{-2\,a} = \frac{1}{\Gamma(a)}\,\int_0^\infty dt\,
t^{a-1}\, e^{-x^2\,t}\qquad,\qquad \Re (a), \Re (x^2)>0\quad,
\eeq
we get
\bea
c_{2a} &=& \int d^dx\, e^{-i\, x_\mu{k^\mu}/k}\,x^{-2\,a} 
= \frac{1}{\Gamma(a)}\,\int_0^\infty dt\,t^{a-1}\, \int d^dx\, e^{-x^2\,t-i\, x_\mu k^\mu/k}\cr
&=&\frac{\pi^\frac{d}{2}}{\Gamma(a)}\,\int_0^\infty dt\,t^{a-\frac{d}{2}-1}\, e^{-\frac{1}{4\,t}}
=(4\pi)^\frac{d}{2}\,\frac{\Gamma(\frac{d}{2}-a)}
{2^{2\,a}\,\Gamma(a)}\quad.
\eea

\noi
A useful relation that follows is (for $d=4$)
\bea\label{eq:c2a/c2a+2m}
\frac{2^{-2\,m}\,c_{2a}}{c_{2a+2m}}
&=& 
\frac{\Gamma(2-a)}{\Gamma(2-a-m)}\,
\frac{\Gamma(a+m)}{\Gamma(a)}
= (-)^m\,(a-1)_m\,(a)_m\cr
&=& (-)^m\,(a-1)\,(a-1+m)\,
\prod_{l=1}^{m-1} (a-1+l)^2\quad.
\eea

So, from \eqref{eq:GmGiorgio} we get:
\bigskip

\noindent\underline{Zero order}
\beq
 G_2^{(0)}(k)= f_{2\,\Delta}(k) = c_{2\Delta}\,k^{2\,\Delta-d}=
\pi^\frac{d}{2}\,\frac{\Gamma(-\nu)}{\Gamma(\Delta)}\,
\left(\frac{k}{2}\right)^{2\,\nu}\quad.
\eeq

Note the different normalization of \eqref{eq:GmGiorgio} 
with respect to the one used in the paper, equation \eqref{eq:propT=0}. 
\bigskip

\noindent\underline{First order}
\bea\label{eq:G1}
 G_2^{(1)}(k)
&=& \frac{1}{ k^4}\,\frac{2\,\Delta\,k^4}{15\,2^4}\,
\left(-f_{2\,\nu}(k)-4\,\partial_\omega^2f_{2\,\nu+2}(k)
\right)\cr
&=& \frac{1}{ k^4}\,
\frac{2\,\Delta\,k^{2\nu}}{15\,2^4}\,
\left(-c_{2\,\nu}-8\,(\nu-1)\,c_{2\,\nu+2}\,
\left(1+2\,(\nu-2)\,\left(\frac{\omega}{k}\right)^2\right)\right)\cr
&=& \frac{ G_2^{(0)}(k)}{ k^4}\,
\frac{2\,\Delta}{15\,2^4}\,
\frac{c_{2\,\nu}}{c_{2\,\nu+4}}\,
\left(-1-8\,(\nu-1)\,\frac{c_{2\,\nu+2}}{c_{2\,\nu}}\,
\left(1+2\,(\nu-2)\,\left(\frac{\omega}{k}\right)^2\right)\right)\cr
&=& \frac{ G_2^{(0)}(k)}{ k^4}\,
\frac{2}{15}\,
\nu\,(\nu^2-1)\,(\nu^2-4)\,C_{2}^{(1)}\left(\frac{\omega}{k}\right)\quad,
\eea
where we have used \eqref{eq:c2a/c2a+2m} and the relation
\beq
\int d^4x\, e^{-i\,x_\mu k^\mu/ x}\,x^{-2\,a}\,\eta^m = 
i^m\,c_{2a+m}\,k^{4-2a}\,\partial_\omega^m k^{2a-4+m}\quad.
\eeq

\bigskip

\noindent\underline{Second order}
\bea\label{eq:orden2}
 G_2^{(2)}(k)
&=& \frac{1}{(kz_h)^8}\,
\frac{2\,\Delta^2\,k^8}{15^2\,2^8}\,
\int d^4x\, e^{-i\,k\cdot x}\,x^{-2\,\Delta +8}\,
\left(C_0^{(1)}(\eta)+C_2^{(1)}(\eta)
+C_4^{(1)}(\eta)\right)\cr
&=& \frac{ G_2^{(0)}(k)}{(kz_h)^8}\,
\frac{2\,\Delta^2}{15^2\,2^8}\,
\frac{c_{2\nu-4}}{c_{2\nu+4}}
\sum_{l=0}^2 I_{2\nu}^{(2l)}(\omega/k)\quad,
\eea
where
\beq
I_{2\nu}^{(2l)}(\omega/k)\equiv \frac{k^{8-2\nu}}{c_{2\nu-4}}\,
\int d^4x\, e^{-i\,k\cdot x}\,x^{-2\,\nu +4}\,
C_{2\,l}^{(1)}(\eta)\quad.
\eeq
We get
\bea
I_{2\nu}^{(0)}(x)&=& 1\quad,\cr
I_{2\,\nu}^{(2)}(x)&=& \frac{\nu-4}{\nu-2}\,
C_{2}^{(1)}(x)\quad,\cr
I_{2\,\nu}^{(4)}(x)&=& \frac{(\nu-4)\,(\nu-5)}{(\nu-1)\,(\nu-2)}\,
C_{4}^{(1)}(x)\quad.
\eea
From \eqref{eq:orden2} we obtain

\bea\label{eq:orden2bis}
 G_2^{(2)}(k)
&=& \frac{ G_2^{(0)}(k)}{( kz_h)^8}\,
\frac{2}{15^2}\,(\nu+2)^2\,(\nu-3)\,(\nu+1)\,
\nu^2\,(\nu-1)^2\,(\nu-2)^2\,
\left(
1 + \frac{\nu-4}{\nu-2}\,
C_{2}^{(1)}(\omega/k)
\right.\cr
&+&\left. 
\frac{(\nu-4)\,(\nu-5)}{(\nu-1)\,(\nu-2)}\,
C_{4}^{(1)}(\omega/k)
\right)\quad.
\eea
For large $\nu$ we have
\bea\label{eq:orden2bislargenu}
\left. \frac{ (kz_h)^8}{ G_2^{(0)}(k)}\, G_2^{(2)}(k)\right|_{\nu\rightarrow\infty}
&=&\frac{2}{15^2}\,\nu^{10}\,
\left(
1 + C_{2}^{(1)}(\omega/k)+ C_{4}^{(1)}(\omega/k)
\right)\non\\
&=& \frac{2}{15^2}\,\nu^{10}\,
\left(4\,\left(\frac{\omega}{k}\right)^2-1\right)^2\quad.
\eea


\begin{thebibliography}{99}

\bibitem{Policastro:2001yb}
G.~Policastro and A.~Starinets,
``On the Absorption by Near Extremal Black Branes,''
Nucl. Phys. B \textbf{610} (2001), 117-143
[arXiv:hep-th/0104065 [hep-th]].

\bibitem{Maldacena:1997re}
J.~M.~Maldacena,
``The Large $N$ Limit of Superconformal Field Theories and Supergravity,''
Adv. Theor. Math. Phys. \textbf{2} (1998), 231-252
[arXiv:hep-th/9711200 [hep-th]].

\bibitem{Gubser:1998bc}
S.~S.~Gubser, I.~R.~Klebanov and A.~M.~Polyakov,
``Gauge Theory Correlators from Noncritical String Theory,''
Phys. Lett. B \textbf{428} (1998), 105-114
[arXiv:hep-th/9802109 [hep-th]].

\bibitem{Witten:1998qj}
E.~Witten,
``Anti-de~Sitter Space and Holography,''
Adv. Theor. Math. Phys. \textbf{2} (1998), 253-291
[arXiv:hep-th/9802150 [hep-th]].


\bibitem{Fitzpatrick:2019zqz}
A.~L.~Fitzpatrick and K.~W.~Huang,
``Universal Lowest-Twist in CFTs from Holography,''
JHEP \textbf{08} (2019), 138
[arXiv:1903.05306 [hep-th]].

\bibitem{Rodriguez-Gomez:2021pfh}
D.~Rodriguez-G\'omez and J.~G.~Russo,
``Correlation Functions in Finite Temperature CFT and Black Hole Singularities,''
JHEP \textbf{06} (2021), 048
[arXiv:2102.11891 [hep-th]].

\bibitem{Rodriguez-Gomez:2021mkk}
D.~Rodriguez-G\'omez and J.~G.~Russo,
``Thermal Correlation Functions in CFT and Factorization,''
JHEP \textbf{11} (2021), 049
[arXiv:2105.13909 [hep-th]].

\bibitem{Parisini:2022wkb}
E.~Parisini, K.~Skenderis and B.~Withers,
``An Embedding Formalism for CFTs in General States on Curved Backgrounds,''
[arXiv:2209.09250 [hep-th]].
	
\bibitem{CarneirodaCunha:2015qln}
B.~Carneiro da Cunha and F.~Novaes,
``Kerr-De Sitter Greybody Factors via Isomonodromy,''
Phys. Rev. D \textbf{93} (2016) no.2, 024045
[arXiv:1508.04046 [hep-th]].

\bibitem{Bonelli:2022ten}
G.~Bonelli, C.~Iossa, D.~P.~Lichtig and A.~Tanzini,
``Irregular Liouville Correlators and Connection Formulae for Heun Functions,''
[arXiv:2201.04491 [hep-th]].

\bibitem{Dodelson:2022yvn}
M.~Dodelson, A.~Grassi, C.~Iossa, D.~Panea Lichtig and A.~Zhiboedov,
``Holographic Thermal Correlators from Supersymmetric Instantons,''
[arXiv:2206.07720 [hep-th]].

\bibitem{Bhatta:2022wga}
A.~Bhatta and T.~Mandal,
``Exact Thermal Correlators of Holographic $CFT$s,''
[arXiv:2211.02449 [hep-th]].


\bibitem{Bajc:2012vk}
B.~Bajc, A.~R.~Lugo and M.~B.~Sturla,
``Spontaneous breaking of a discrete symmetry and holography,''
JHEP \textbf{04} (2012), 119
[arXiv:1203.2636 [hep-th]].

\bibitem{Aharony:1999ti}
O.~Aharony, S.~S.~Gubser, J.~M.~Maldacena, H.~Ooguri and Y.~Oz,
``Large $N$ Field Theories, String Theory and Gravity,''
Phys. Rept. \textbf{323} (2000), 183-386
[arXiv:hep-th/9905111 [hep-th]].

\bibitem{Grozdanov:2019uhi}
S.~Grozdanov, P.~K.~Kovtun, A.~O.~Starinets and P.~Tadi\'c,
``The Complex Life of Hydrodynamic Modes,''
JHEP \textbf{11} (2019), 097
[arXiv:1904.12862 [hep-th]].

\bibitem{Tricomi:1951}
F. Tricomi and A. Erd\'elyi,
``The asymptotic expansion of a ratio of gamma functions,''
Pacific Journal of Mathematics Vol. 1 (1951), No. 1, 133-142.


\bibitem{NIST:DLMF}
NIST Handbook of Mathematical Functions, 
Edited by Frank W. J. Olver, 
Daniel W. Lozier, Ronald F. Boisvert and
Charles W. Clark,
National Institute of Standards and Technology, Maryland and University of Maryland, 
Chapter 10, https://dlmf.nist.gov/10.41.

\bibitem{Langer}
R. Langer, “On the asymptotic solutions of ordinary differential equations, with reference to the Stokes’ 
phenomenon about a singular point”, Trans. Amer. Math. Soc. {\bf 37} no. 3, 397–416 (1935).

\bibitem{Olver}
F. W. J. Olver, “The asymptotic solution of linear differential equations of the second order in a domain 
containing one transition point”, Philos. Trans. Roy. Soc. London. Ser. A. {\bf 249}, 65–97 (1956), “Uniform 
asymptotic expansions of solutions of linear second-order differential equations for large values of a parameter”, 
Philos. Trans. Roy. Soc. London. Ser. A {\bf 250} 479–517 (1958).


\bibitem{gradshteyn2007}
I.S. Gradshteyn and I.M. Ryzhik., 
``Table of integrals, series, and products,'' 
Elsevier/Academic Press, Amsterdam (2007).



\end{thebibliography}
\end{document}